\documentclass[lettersize,journal]{IEEEtran}
\usepackage{amsmath,amsfonts}
\usepackage{graphicx}
\usepackage{amsmath}
\usepackage{amssymb}
\usepackage{array}
\usepackage{subfig}
\usepackage{textcomp}
\usepackage{stfloats}
\usepackage{url}
\usepackage{verbatim}
\usepackage{cite}

\newtheorem{definition}{\textbf{Definition}}
\usepackage[ruled,vlined]{algorithm2e}
\usepackage[table,xcdraw]{xcolor}
\usepackage{multirow}

\begin{document}

\title{Really Unlearned? Verifying Machine Unlearning via Influential Sample Pairs}

\author{Heng Xu, Tianqing Zhu*, Lefeng Zhang, Wanlei Zhou
\thanks{
*Tianqing Zhu is the corresponding author. 
Heng Xu and Tianqing Zhu are with the Center for Cyber
Security and Privacy and the School of Computer Science, University of Technology Sydney, Ultimo, NSW 2007, Australia (e-mail: heng.xu-2@student.uts.edu.au; tianqing.zhu@uts.edu.au).
Lefeng Zhang and Wanlei Zhou are with the City University of Macau, Macau (e-mail: lfzhang@cityu.edu.mo; wlzhou@cityu.edu.mo). 
}
}

\markboth{Journal of \LaTeX\ Class Files,~Vol.~14, No.~8, August~2021}%
{Shell \MakeLowercase{\textit{et al.}}: A Sample Article Using IEEEtran.cls for IEEE Journals}


\maketitle

\begin{abstract}
Machine unlearning enables pre-trained models to eliminate the effects of partial training samples. Previous research has mainly focused on proposing efficient unlearning strategies. However, the verification of machine unlearning, or in other words, how to guarantee that a sample has been successfully unlearned, has been overlooked for a long time. Existing verification schemes typically rely on machine learning attack techniques, such as backdoor or membership inference attacks. As these techniques are not formally designed for verification, they are easily bypassed when an untrustworthy MLaaS undergoes rapid fine-tuning to merely meet the verification conditions, rather than executing \emph{real} unlearning. In this paper, we propose a formal verification scheme, IndirectVerify, to determine whether unlearning requests have been successfully executed. We design influential sample pairs: one referred to as \emph{trigger samples} and the other as \emph{reaction samples}. Users send unlearning requests regarding trigger samples and use reaction samples to verify if the unlearning operation has been successfully carried out. We propose a perturbation-based scheme to generate those influential sample pairs. The objective is to perturb only a small fraction of trigger samples, leading to the reclassification of reaction samples. This indirect influence will be used for our verification purposes. In contrast to existing schemes that employ the same samples for all processes, our scheme, IndirectVerify, provides enhanced robustness, making it less susceptible to bypassing processes.
\end{abstract}


\section{Introduction}
\IEEEPARstart{M}{achine} learning as a service (MLaaS) is gaining increasing popularity in machine learning and cloud computing~\cite{DBLP:journals/tdsc/WengWCHW22}. 
It enables users to delegate complex training tasks to a model provider, accessing the model in a black-box manner, thereby enhancing accessibility and efficiency in machine learning deployment. However, similar to traditional machine learning models, MLaaS have also raised various privacy and security concerns, as they can potentially memorize sensitive information from the training dataset.~\cite{DBLP:journals/csur/HuSSDYZ22,DBLP:journals/tifs/ZhuYZLZ23}. Machine unlearning has attracted attention recently as a solution. It refers to the process of removing the effect of partial training samples from models~\cite{DBLP:journals/csur/XuZZZY24,DBLP:journals/tifs/ZhangZZXZ23}. This technological advancement has received significant attention recently due to various factors. For example, legislators around the world have introduced laws and regulations that grant users \textit{ the right to be forgotten}, which requires companies to remove user data from their systems after receiving unlearning requests~\cite{webpage:GDPR}.

Previous machine unlearning studies have primarily focused on efficiently unlearning samples from models~\cite{DBLP:conf/sp/CaoY15,DBLP:conf/sp/BourtouleCCJTZL21,10.1145/3485447.3511997,DBLP:conf/ccs/Chen000H022,DBLP:conf/aaai/GravesNG21}. For example, Bourtoule et al.~\cite{DBLP:conf/sp/BourtouleCCJTZL21} proposed a ``sharded, isolated, sliced and aggregated'' ($\mathbf{SISA}$) unlearning framework, inspired by the data segmentation technique. Guo et al.~\cite{DBLP:conf/icml/GuoGHM20} calculated the effect of samples that need to be unlearned and directly adjusted model parameters to compensate for those effects. However, those schemes usually assume that model providers in MLaaS are trustworthy and will honestly unlearn samples based on received unlearning requests~\cite{DBLP:conf/eccv/GolatkarAS20,DBLP:conf/cvpr/GolatkarARPS21,DBLP:conf/cvpr/GolatkarAS20,DBLP:conf/infocom/LiuXYWL22}. In practice, model providers may not be entirely trustworthy due to various reasons. For example, they may not comply with unlearning requests due to the time-consuming nature of unlearning operations~\cite{TIFSYu2024,DBLP:journals/popets/SommerSWM22}. In this situation, an efficient scheme to verify unlearning is urgently needed to confirm that model providers indeed unlearn the samples requested by users. In addition to this, reasonable verification schemes can also provide model providers with self-verification that they have actually performed the unlearning operation.

Only a limited number of studies have been proposed to address these issues. Weng et al.\cite{DBLP:journals/corr/abs-2210-11334} suggested employing trusted hardware to enforce proof of unlearning but relied on the trusted execution environments on MLaaS. David et al.\cite{DBLP:journals/popets/SommerSWM22} and Yu et al.~\cite{TIFSYu2024} proposed verification schemes based on backdoor attacks, in order to assess if the model provider successfully performs unlearning by evaluating the success rate of the backdoor before and after the unlearning request. Other schemes are based on the difference between the outputs of training and test samples, and combined with membership inference attacks~(MIAs) for verification~\cite{DBLP:journals/tifs/LiuXMW22}. 

However, those verification schemes did not consider the possibility that model providers may bypass those verification schemes through a rapid fine-tuning process. In the case of backdoor-based verification schemes, model providers can quickly fine-tune a model using unlearning samples with true labels. This process alters the output of those samples, thereby affecting the results of the verification. Similarly, for MIAs-based verification schemes, model providers can bypass it by aligning the outputs of unlearning samples with those of corresponding test samples. Therefore, there is an urgent need for an effective and robust formal verification solution.

In this paper, we propose a novel sample-pair verification scheme, IndirectVerify, to validate the machine unlearning process in MLaaS. First, to show the existing bypass problems, we propose two methods to bypass existing backdoor and MIA-based verification schemes. This highlights the unreliability of current verification methods when faced with untrustworthy MLaaS. Second, we propose IndirectVerify, which mainly involves constructing influential sample pairs. This pair includes trigger samples and a reaction sample, where the presence or absence of trigger samples within the training process will affect the classification results of the reaction sample. Users can inject trigger samples into the training dataset to induce misclassification of reaction samples. When users request the unlearning of trigger samples; if executed successfully, the model will alter the classification result of the reaction samples. This change in classification result will be used for our verification. 


Specifically, we propose an optimization-based method for generating trigger samples that lead to the misclassification of selected reaction samples. To maintain model performance after embedding trigger samples for verification, we transform our perturbation objective to a gradient matching problem~\cite{DBLP:conf/iclr/GeipingFHCT0G21}. The objective is to modify those trigger samples and align their gradients with those of the targeted reaction samples. This alignment is crucial for our verification purpose, as minimizing the loss on modified trigger samples simultaneously reduces the loss on the targeted reaction samples. This will serve our purpose of classifying reaction samples as our target label when trigger samples are in the training set. We provide theoretical proofs and analyses of IndirectVerify based on influence functions, which illustrate how unlearning each training sample affects reaction samples, providing a theoretical foundation for IndirectVerify. We also briefly emphasize the robustness of IndirectVerify by explaining how it resists bypassing methods.

In summary, we make the following contributions.

\begin{itemize}
    \item We take the first step in addressing the machine unlearning verification problem when confronted with a potent, untrusted MLaaS. To prove that, we propose fine-tuning-based methods to bypass existing verification schemes.
    \item We propose a perturbation-based verification scheme named IndirectVerify. It involves two samples: trigger samples and a reaction sample. Trigger samples serve as the sample in the unlearning request, and reaction samples are used for subsequent verification. The presence or absence of trigger samples can affect reaction samples.
    \item We provide theoretical proofs and analyses of IndirectVerify based on influence functions to illustrate the effectiveness of our scheme. We also explain its robustness by explaining how it resists fine-tuning-based bypassing methods.
    \item We comprehensively evaluate IndirectVerify, considering various perspectives, datasets, and models. The results demonstrate that IndirectVerify maintains model utility while achieving robust verification effectiveness.
\end{itemize}

\section{Preliminary}
\subsection{Related Works}
\subsubsection{Machine Unlearning}

The machine learning community has proposed numerous unlearning schemes in response to \textit{the right to be forgotten}. In our previously published survey paper~\cite{DBLP:journals/csur/XuZZZY24}, we conducted a comprehensive survey encompassing recent studies on machine unlearning. This survey thoroughly summarized several crucial facets, including: (i) the motivation behind machine unlearning; (ii) the objectives and desired outcomes associated with the unlearning process; (iii) a novel taxonomy for systematically categorizing existing machine unlearning schemes according to their rationale and strategies; and (iv) the characteristics as well as the advantages and disadvantages of existing verification schemes.

In general, existing unlearning schemes are mainly based on \textit{data reorganization} and \textit{model manipulation} techniques~\cite{DBLP:journals/csur/XuZZZY24}. Data reorganization refers to the technique by which model providers reorganize the training dataset to accelerate the unlearning  process~\cite{DBLP:conf/sp/CaoY15,DBLP:conf/sp/BourtouleCCJTZL21,10.1145/3485447.3511997,DBLP:conf/ccs/Chen000H022,DBLP:conf/aaai/GravesNG21}. Bourtoule et al.~\cite{DBLP:conf/sp/BourtouleCCJTZL21} presented a $\mathbf{SISA}$ architecture, similar to the existing ensemble training methodologies, to achieve the unlearning purpose. Other similar schemes are used in recommendation tasks~\cite{10.1145/3485447.3511997} and graph data classification tasks~\cite{DBLP:conf/ccs/Chen000H022}. Graves et al.~\cite{DBLP:conf/aaai/GravesNG21} presented a framework that leverages random relabeling and fine-tuning strategies to facilitate class-level unlearning. For the unlearning schemes based on model manipulation, the model provider aims to realize unlearning operations by directly adjusting the model's parameters. Guo et al.~\cite{DBLP:conf/icml/GuoGHM20} proposed a certified removal unlearning scheme based on influence function~\cite{DBLP:conf/icml/KohL17} and differential privacy~\cite{DBLP:journals/tkde/ZhuLZY17}. In~\cite{DBLP:conf/eccv/GolatkarAS20,DBLP:conf/cvpr/GolatkarARPS21,DBLP:conf/cvpr/GolatkarAS20,DBLP:conf/infocom/LiuXYWL22}, similar approaches were proposed, which typically involve directly removing the influence of the corresponding samples from the model. ~\cite{DBLP:conf/iwqos/LiuMYWL21,DBLP:conf/www/Wang0XQ22} considered the unlearning requests in the federated learning setting and proposed model pruning-based methods. Wu et al.~\cite{DBLP:conf/icml/WuDD20} and Brophy et al.~\cite{DBLP:conf/icml/BrophyL21} proposed parameter replacement-based unlearning schemes that involve precomputing and subsequently replacing operations.

\subsubsection{Unlearning Verification}
Currently, there are no specialized schemes designed for verifying machine unlearning. The methods employed for this purpose are usually drawn from various other fields. Generally, existing methods can be broadly classified into the following two categories: \textit{empirical evaluation} and \textit{theoretical analysis}.

Empirical evaluation schemes usually use attack methods to evaluate how much information about samples that needs to be unlearned remains within the model. Model inversion attack~\cite{DBLP:conf/ccs/FredriksonJR15} has been used in~\cite{DBLP:conf/aaai/GravesNG21}, while membership inference attacks~(MIAs) has been used in ~\cite{DBLP:conf/aaai/GravesNG21,DBLP:conf/iwqos/LiuMYWL21,DBLP:conf/eccv/GolatkarAS20,DBLP:conf/cvpr/GolatkarARPS21}. Cao et al.~\cite{DBLP:conf/sp/CaoY15} performed data pollution attacks to verify whether the model's performance, post-unlearning process, was restored to its initial state. Similar schemes are adopted in~\cite{TIFSYu2024,DBLP:journals/popets/SommerSWM22} based on the backdoor attack in MLaaS. However, those schemes, such as model inversion attack~\cite{DBLP:conf/aaai/GravesNG21},  MIAs~\cite{DBLP:conf/aaai/GravesNG21,DBLP:conf/iwqos/LiuMYWL21,DBLP:conf/eccv/GolatkarAS20,DBLP:conf/cvpr/GolatkarARPS21}, backdoor~\cite{TIFSYu2024,DBLP:journals/popets/SommerSWM22} can be bypassed by malicious MLaaS. This is because the samples used in the unlearning request are the same as those used for verification. Malicious MLaaS can bypass the verification by modifying the output of the samples directly to meet the verification condition, rather than indeed executing the unlearning process. We will discuss this in detail in Section~\ref{sec:bypass}.

In addition to attack-based methods, Brophy et al.~\cite{DBLP:conf/icml/BrophyL21}, Wang et al.~\cite{DBLP:conf/www/Wang0XQ22}, Liu et al.~\cite{DBLP:conf/infocom/LiuXYWL22},  Wu et al.~\cite{DBLP:conf/icml/WuDD20} and Golatkar et al.~\cite{DBLP:conf/eccv/GolatkarAS20,DBLP:conf/cvpr/GolatkarAS20,DBLP:conf/cvpr/GolatkarARPS21} used accuracy-based verification schemes. However, the essential purpose of machine unlearning is to remove the effects of partial samples from the model. This process is similar to removing those samples from the training dataset to the test dataset. Verification schemes based on accuracy cannot reflect whether samples have been unlearned. There is no direct relationship between the unlearned samples' accuracy and the unlearning scheme's effectiveness~\cite{DBLP:conf/icml/BrophyL21,DBLP:conf/www/Wang0XQ22,DBLP:conf/infocom/LiuXYWL22,DBLP:conf/icml/WuDD20,DBLP:conf/eccv/GolatkarAS20,DBLP:conf/cvpr/GolatkarAS20,DBLP:conf/cvpr/GolatkarARPS21}. Baumhauer et al.~\cite{DBLP:conf/sp/BourtouleCCJTZL21}, Golatkar et al.~\cite{DBLP:conf/cvpr/GolatkarAS20,DBLP:conf/cvpr/GolatkarARPS21} and Liu et al.~\cite{DBLP:conf/iwqos/LiuMYWL21} measured the similarity of the resulting distributions of pre-softmax outputs. There are also some efforts to measure the similarity between the parameter distributions of the model after unlearning and the model after retraining from scratch~\cite{DBLP:conf/infocom/LiuXYWL22}. However, those methods have been proven to be ineffective~\cite{DBLP:conf/uss/ThudiJSP22}. Theoretical analysis schemes typically ensure that the unlearning operation can indeed unlearn sample information~\cite{DBLP:conf/icml/GuoGHM20}. However, these approaches often come with specific unlearning strategies and are ineffective when dealing with complex models and large datasets~\cite{DBLP:conf/nips/SekhariAKS21}. 

\subsubsection{Discussion of Related Works}
The limitations of existing machine unlearning verification schemes can be summarized as follows: (1). Attack-based schemes, such as backdoor attacks~\cite{TIFSYu2024,DBLP:journals/popets/SommerSWM22} and membership inference attacks~\cite{DBLP:conf/aaai/GravesNG21,DBLP:conf/iwqos/LiuMYWL21,DBLP:conf/eccv/GolatkarAS20,DBLP:conf/cvpr/GolatkarARPS21}, are susceptible to malicious MLaaS. MLaaS can bypass those verification schemes based on fine-tuning processes; (2). Accuracy-based verification schemes usually fail to reflect the effectiveness of unlearning in removing sample effects~\cite{DBLP:conf/icml/BrophyL21,DBLP:conf/www/Wang0XQ22,DBLP:conf/infocom/LiuXYWL22,DBLP:conf/icml/WuDD20,DBLP:conf/eccv/GolatkarAS20,DBLP:conf/cvpr/GolatkarAS20,DBLP:conf/cvpr/GolatkarARPS21}. (3). Distribution similarity-based verification schemes~\cite{DBLP:conf/sp/BourtouleCCJTZL21,DBLP:conf/cvpr/GolatkarAS20,DBLP:conf/cvpr/GolatkarARPS21,DBLP:conf/iwqos/LiuMYWL21}, like output distributions and parameter distributions, have been demonstrated to be ineffective~\cite{DBLP:conf/uss/ThudiJSP22}. (4). Theoretical analysis schemes~\cite{DBLP:conf/icml/GuoGHM20,DBLP:conf/nips/SekhariAKS21}, while ensuring the unlearning results, are often limited by specific unlearning strategies and show inefficacy with complex models and large datasets. Consequently, there is a need for mechanisms through which users can verify whether model providers comply with their unlearning requests.

\begin{table}
  \caption{Notations}
  \renewcommand{\arraystretch}{1.2}
  \label{tab:notations}
  \centering
  \begin{tabular}{c|c}
    \hline
    Notations &  Explanation \\
    \hline
    $\mathcal{D}$                       &The training dataset\\
    $\left( \mathbf{x}, y \right)$      &One instance in $\mathcal{D}$\\
    $M$                                 &The original trained model\\
    $\mathcal{A}(\cdot)$                &The machine learning process\\
    $\mathcal{D}_{u}$                   &The unlearning dataset for unlearning process\\
    $\mathcal{D}_{r}$                   &The remaining dataset after unlearning process\\
    $\mathcal{U}(\cdot)$                &The machine unlearning process\\
    $M_{u}$                             &The model after machine unlearning process\\
    $\mathcal{V}(\cdot)$                &The verification process for machine unlearning\\
    $O(\cdot)$                          &The outputs of the model\\
    $\ell \left(M\left(\mathbf{x}\right), y\right)$     &The loss of one instance\\
    $R(\theta)$                         &The empirical risk\\
    $\hat{\theta}$                      &The empirical risk minimizer\\
    $\mathbf{x}_{t}$                    &The test sample used for verification\\
    $\Delta \ell$                       &The change of model loss\\
    $\eta(\cdot)$                       &The query request\\
    $\upsilon(\cdot)$                   &The unlearning request\\
    \hline
	\end{tabular}
\end{table}

\subsection{Machine Unlearning and MLaaS}
MLaaS is a comprehensive framework that typically comprises two key entities: \textit{data provider} and \textit{model provider}. Data provider uploads their own data to model providers for model training. We use $\mathcal{D}=\left\{\left(\mathbf{x}_{1}, y_{1}\right), \left(\mathbf{x}_{2}, y_{2}\right),...,\left(\mathbf{x}_{n}, y_{n}\right) \right\} \subseteq \mathbb{R}^{d} \times \mathbb{R}$ to represent the dataset of a data provider, in which each sample $\mathbf{x}_{i} \in \mathcal{X}$ is a $d$-dimensional vector, $y_{i} \in \mathcal{Y}$ is the corresponding label, and $n$ is the size of $\mathcal{D}$. Let $\mathcal{A}$ be a (randomized) learning algorithm that trains on $\mathcal{D}$ and outputs a model $M$. $M = \mathcal{A}(\mathcal{D})$, $M \in \mathcal{H}$, where $\mathcal{H}$ is the hypothesis space. 
Data providers sometimes wish to remove their partial samples from the trained model and send an unlearning request to the model provider. Let $\mathcal{D}_{u} \subset \mathcal{D}$ be a subset of the training dataset, whose influence one data provider wants to remove. Let its complement $\mathcal{D}_{r}=\mathcal{D}_{u}^{\complement} = \mathcal{D}/\mathcal{D}_{u}$ be the remaining dataset. Now, we give the definition of machine unlearning.

\begin{definition}[Machine Unlearning~\cite{DBLP:conf/sp/CaoY15}]
    \label{Definition:Machineunlearning}
    Consider a cluster of samples that one data provider wants to remove from the already-trained model, denoted as $\mathcal{D}_{u}$. An unlearning process $\mathcal{U}(M, \mathcal{D}, \mathcal{D}_{u})$ is defined as a function from an already-trained model $M = \mathcal{A}(\mathcal{D})$, a training dataset $\mathcal{D}$, and an unlearning dataset $\mathcal{D}_{u}$ to a model $M_u$, which ensures that the unlearned model $M_{u}$ performs as though it had never seen $\mathcal{D}_{u}$.
\end{definition}

After machine unlearning process, a verification process $\mathcal{V}(\cdot)$ can make a distinguishable check, that is, $\mathcal{V}\left(M, \mathcal{D}_u\right) \neq \mathcal{V}\left(M_u,\mathcal{D}_u\right)$.  $\mathcal{V}(\cdot)$ can be used to measure whether a model provider has indeed unlearned the requested samples $\mathcal{D}_{u}$. Other symbols that appear in this paper and their corresponding descriptions are listed in TABLE~\ref{tab:notations}. 


\section{Problem Defination}
\subsection{Existence Verification Scheme Process}
\begin{figure}
    \centering
    \includegraphics[width=1\linewidth]{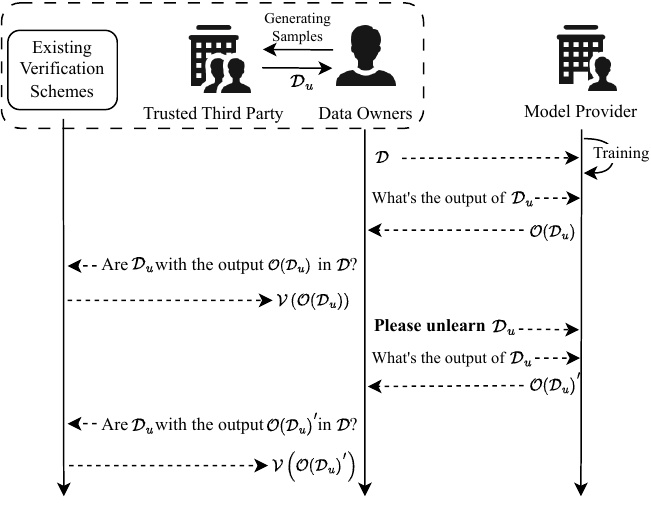}
    \caption{Existing Verification Scheme Process.}
    \label{fig:verificationprocess}
\end{figure}

Existing verification schemes in MLaaS follow the process outlined in Figure~\ref{fig:verificationprocess}. It mainly consists of a series of steps. Firstly, trigger samples $\mathcal{D}_u$ are selected and incorporated into the training dataset. Those trigger samples $\mathcal{D}_u$ are either generated by the data provider or obtained from a trusted third party. The model is then trained based on this aggregated dataset. In the second step, the data provider queries the output of $\mathcal{D}_u$, and the model provider returns the corresponding outputs. A pre-prepared verification scheme is employed by the data provider to determine if $\mathcal{D}_u$ with the received outputs is present in the model. Following this verification, the data provider submits an unlearning request for the $\mathcal{D}_u$. Then, the data provider checks the model's outputs again, using the same verification scheme to assess the presence of $\mathcal{D}_u$. By comparing the results before and after unlearning, the data provider can ascertain whether the model has indeed undergone the unlearning operation.

Taking MIAs-based verification schemes~\cite{DBLP:conf/aaai/GravesNG21,DBLP:conf/iwqos/LiuMYWL21,DBLP:conf/eccv/GolatkarAS20,DBLP:conf/cvpr/GolatkarARPS21} and backdoor-based verification schemes~\cite{TIFSYu2024,DBLP:journals/popets/SommerSWM22} as examples, if $\mathcal{D}_u$ that need to be unlearned are verified as existing in $M$ but not $M_u$, that is $\mathcal{V}\left(O(\mathcal{D}_{u})\right) = true$ and $\mathcal{V}(O(\mathcal{D}_{u})^{'}) = false$, the data provider can conclude that the model provider has performed the unlearning operation.

\subsection{Case Studies: the bypass of current verification schemes}
\label{sec:bypass}

Currently, there are two main techniques used for verifying the unlearning process in MLaaS,  namely membership inference attacks (MIAs)~\cite{DBLP:journals/tifs/LiuXMW22,DBLP:conf/aaai/GravesNG21,DBLP:conf/iwqos/LiuMYWL21,DBLP:conf/eccv/GolatkarAS20,DBLP:conf/cvpr/GolatkarARPS21} and backdoor attacks~\cite{TIFSYu2024,DBLP:journals/popets/SommerSWM22}. Membership inference attacks is a type of privacy attack that targets machine learning models. It usually requires pre-training an attack model based on the output of one dataset that closely resembles the training dataset as much as possible. However, in MLaaS, fulfilling this condition becomes challenging as it is impractical for data providers to obtain the output of the entire training dataset. Moreover, since the performance of MIAs is usually significantly influenced by the type of dataset and the configuration of the attack model, they usually achieve less-than-ideal success rates. These characteristics make MIAs not entirely suitable for verifying the unlearning process. Backdoor attacks involve manipulating a model during the training phase. Verification schemes based on backdoor attacks also inherit the drawbacks of the backdoor itself. Backdoors pose security risks to trained models, which will hinder the widespread adoption for verification in MLaaS. 

In addition to the above issues, the samples used for verification and those used for unlearning requests are the same in the above two schemes. This gives the model provider a chance to know what the model output will be after the unlearning operation. Based on this, they can specify the output of the samples that are used for verification, thus bypassing the above schemes.
Here, we illustrate how to bypass the MIAs-based verification schemes~\cite{DBLP:conf/aaai/GravesNG21,DBLP:conf/iwqos/LiuMYWL21,DBLP:conf/eccv/GolatkarAS20,DBLP:conf/cvpr/GolatkarARPS21} and backdoor-based verification schemes mentioned in~\cite{TIFSYu2024,DBLP:journals/popets/SommerSWM22}. 

\subsubsection{Bypassing MIAs-based Schemes}
MIAs-based verification schemes~\cite{DBLP:conf/aaai/GravesNG21,DBLP:conf/iwqos/LiuMYWL21,DBLP:conf/eccv/GolatkarAS20,DBLP:conf/cvpr/GolatkarARPS21} determine whether given samples have been unlearned based on the model's output. If the model's output matches the output from the training dataset, MIAs conclude that those samples have not been unlearned. On the other hand, if the sample's output matches that of the test dataset, MIAs conclude that the sample has been unlearned. Based on this, after receiving the unlearning request, the model provider can make the output $\mathcal{O}(\mathcal{D}_{u})$ of received samples consistent with a sample from the test dataset $\mathcal{O}(\mathcal{D}_{a})$, thus bypassing MIAs-based schemes. Specifically, they can fine-tune the model using the following loss rather than executing the unlearning process:

\begin{equation}
    L = \frac{1}{n}\sum_{i=1}^{n} \ell \left(M\left(\mathbf{x}_i\right), y_i\right) + \lambda ||O(\mathcal{D}_u)-O(\mathcal{D}_a)||_{2}
    \label{equation:attack_loss}
\end{equation}

The front part of Equation~\ref{equation:attack_loss} is the original training loss, and the latter part uses $l_2$ norm to encourage the similarity between the two outputs. $O(\mathcal{D}_u)$ represents the outputs of samples that need to be unlearned, while $O(\mathcal{D}_a)$ denotes the outputs of one sample $\mathcal{D}_a$ from the test dataset with the same label as $\mathcal{D}_u$.

\subsubsection{Bypassing Backdoor-based Schemes}

For the backdoor-based verification schemes mentioned in~\cite{TIFSYu2024,DBLP:journals/popets/SommerSWM22}, the model provider can use existing backdoor removal schemes~\cite{DBLP:conf/cvpr/MuN0WMJ023} to bypass the verification. Here, we employ a simple yet effective approach that is based on assigning original true labels and fine-tuning model:
\begin{equation}
    L = \frac{1}{n} \sum_{i=1}^{n} \ell \left(M\left(\mathbf{x}_i\right), y_{i}\right)
    \label{equation:attack_loss_for_backdoor_random}
\end{equation}

where $y_{i}$ are the true labels for received unlearning samples $\mathbf{x}_i$. Given that the model provider has access to all the samples from the data provider, it is straightforward for the model provider to obtain any label from the dataset and assign true labels to any received samples. Therefore, the above task is not challenging for the model provider.

Both of the above bypassing schemes are simpler than methods involving unlearning, offering the model provider a way to reduce resource costs and meet verification requirements. Detailed evaluation results will be presented in Section~\ref{sec:experimentbypass}.

\subsection{Objectives}

In this paper, our objective is to address three primary issues associated with machine unlearning verification. First, we aim to propose an efficient verification mechanism that enables the data provider to confirm if the model provider has indeed executed the unlearning process. Second, we need to consider situations in which the model provider may be untrustworthy, potentially bypassing the verification process through easily implementable fine-tuning procedures. For instance, this could involve output alignment or fine-tuning based on the original true labels, as discussed in Section~\ref{sec:bypass}. Third and finally, the scheme we propose must minimize any adverse impact on the model. Formal definitions of these problems are provided below, with key notations summarized in Table~\ref{tab:notations}.

Specifically, the data provider sends its own dataset $\mathcal{D}$ to the model provider in MLaaS, where the model provider trains a model, $M$, using one (randomized) learning algorithm $\mathcal{A}(\cdot)$. Upon completing the training process, and in addition to using the trained model for regular prediction requests, the data provider can also submit unlearning requests, aiming to remove the effect of partial samples from the model. While the model provider faithfully follows the MLaaS specification for its service rewards; however, it might not fully comply with the unlearning requests to save cost~\cite{TIFSYu2024,DBLP:journals/popets/SommerSWM22} or aims to serve other commercial purposes~\cite{DBLP:journals/tifs/LiuXMW22,DBLP:journals/tifs/LiZYJWX23}. Consequently, we aim to provide a method for data providers to verify if their samples have been unlearned from the trained model.

For the purposes of this paper, we assume that data provider can only access the trained model in a black-box manner, with access to only some of the samples for verification. Additionally, data provider also has knowledge of the model architecture employed by its model provider. During both pre- and post-verification stages, the data provider should perform verification operations, either independently or through a trusted third party. This assumption is made without loss of generality, as in many MLaaS scenarios, normal data providers have limited ability to do such a verification process. We also assume that the model provider knows who sent the requests to unlearn which samples and aims to bypass the data provider's verification based on fewer computational resources rather than complying with the unlearning request.

\section{Methodology}
\subsection{Overview}
Model providers can bypass existing verification schemes, since these methods usually use the same samples for both unlearning requests and verification process. This allows the model provider to fulfill the verification conditions based on the samples in the unlearning request by directly altering the output of those samples. In this paper, we propose a novel scheme, called IndirectVerify, that consists of influential sample pairs: one referred to as trigger samples and the other as reaction samples. Users send unlearning requests concerning trigger samples and use reaction samples to verify the success of machine unlearning.

\begin{figure}
    \centering
    \includegraphics[width=1\linewidth]{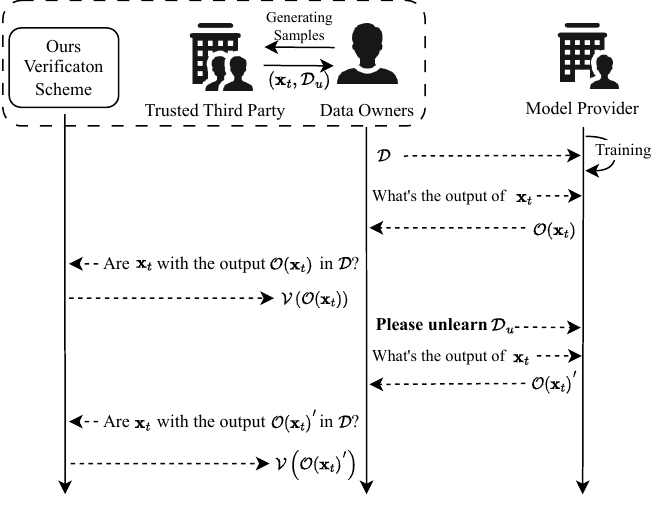}
    \caption{Our Verification Scheme}
    \label{fig:ourscheme}
\end{figure}

Theoretically, the presence or absence of any training sample in the training set can influence a given test sample~\cite{DBLP:conf/icml/KohL17}. Some training samples, when present, may increase the loss of this test sample, while others may decrease the loss. We refer to the former as harmful samples and the latter as helpful samples. The fundamental idea of our scheme, IndirectVerify, is to generate those harmful samples~(trigger samples $\mathcal{D}_u$ in our scheme) for a given test sample~(reaction sample $\mathbf{x}_{t}$ in our scheme), causing the loss of this test sample to become sufficiently large, leading to misclassification. Upon performing the unlearning operation of those harmful samples, the loss of the test sample decreases, allowing it to be reclassified correctly. This before-and-after change serves as the basis of our verification.

As shown in Figure~\ref{fig:ourscheme}, IndirectVerify differs from the one mentioned in Figure~\ref{fig:verificationprocess} mainly in that the samples in unlearning requests and those used for verification are not the same. In Figure~\ref{fig:verificationprocess}, the data provider infers whether the model provider has performed unlearning based on the outputs of the trigger samples. In IndirectVerify, we 
use the reaction samples, rather than trigger samples. To this end, our scheme, IndirectVerify mainly involves the following two steps: \textit{influential sample pair generation} and \textit{unlearning process verification}.

\subsection{Influential Sample Pair Generation}
\label{sec:influential_sample_pair_generation}


This section will introduce how to generate the influential sample pairs $(\mathcal{D}_u,\mathbf{x}_{t})$. We assume that the data provider knows the model architecture used by the model provider. However, it has no knowledge of model training process employed by model provider. This includes how the model is initialized and the hyperparameter settings during training process, such as learning rate and batch size. The data provider can access limited samples and seek to modify these samples. The objective is to induce misclassification of a specific test sample when these changed samples are involved in the training process. Once these changed images are generated, the data provider can upload them to model provider for model training, along with the original training dataset.

Formally, the task of the data provider is as follows:

\begin{equation}
    \begin{aligned}
    &\min \ell \left(M_{\theta}\left(\mathbf{x}_t\right), y_{\text {target}}\right) \quad\\
    &\text { s.t. } \theta \in \underset{\theta}{\arg \min }(\frac{1}{p} \sum_{i=1}^p \ell\left(M\left(\mathbf{x}_i+\Delta_i\right), y_i\right) +\\
    &~~~~~~~~~~~~~\frac{1}{n-p} \sum_{i=1}^{n-p} \ell\left(M\left(\mathbf{x}_i\right), y_i\right))\\
    \end{aligned}
\end{equation}

\begin{algorithm}[!t]
	\small
 	\caption{Generation and Injection}
	\label{algorithm:generation_and_injection}
	\LinesNumbered 
	\KwIn{One text sample $\mathbf{x}_t$ and partial training set $\mathcal{Q}$, where $\mathcal{Q}=\left\{\left(\mathbf{x}_{1}, y_{1}\right), \left(\mathbf{x}_{2}, y_{2}\right),...,\left(\mathbf{x}_{p}, y_{p}\right) \right\}$, $p\ll n$.}
	\KwOut{The final model $M$ and the training completion response.}
        \textbf{\underline{Data provider executes:}} \\
        ~~~~~// \textit{Constructing influential sample pairs}\\
        ~~~~~Modifying $\mathcal{Q}$ to $\mathcal{D}_u$ based on $\mathcal{B}(\Delta, \theta)=1-\frac{\left\langle \Phi, \Psi \right\rangle}{\left\|\Phi\right\| \cdot\left\|\Psi\right\|}$ until results in a misclassification of $\mathbf{x}_t$.\\
        ~~~~~Send dataset $\mathcal{D}_u$ with own dataset to model provider.\\
        ~~~~~Query the classification results $O(\mathbf{x}_{t})$ of $\mathbf{x}_{t}$.\\
        ~~~~~Verify if the sample with output $O(\mathbf{x}_{t})$ is in the model, and return $\mathcal{V}\left(O(\mathbf{x_t})\right)$. \\
        \textbf{\underline{Model provider executes:}} \\
        ~~~~~Model provider train model based on the received dataset.\\
        ~~~~~Send a training completion response to the data provider.\\
        ~~~~~\textbf{Upon} receiving a prediction request for $\mathbf{x}${\\
             ~~~~~~~~~~\Return {$O(\mathbf{x})$}.\\
        }
\end{algorithm}

\begin{algorithm}[!t]
	\small
 	\caption{Re-query and Verification}
	\label{algorithm:verification}
	\LinesNumbered 
	\KwIn{Influential sample pairs $(\mathcal{D}_u, \mathbf{x}_t)$.}
	\KwOut{Whether model provider has unlearned samples $\mathcal{D}_u$}
        \textbf{Data provider executes}: \\
        ~~~~~// \textit{Verification Process.}\\
        ~~~~~Sending the unlearning request regarding samples $\mathcal{D}_u$.\\
        ~~~~~Re-query the classification results $O(\mathbf{x}_{t})^{'}$ of $\mathbf{x}_{t}$.\\
        ~~~~~Re-query the result of output $O(\mathbf{x}_{t})^{'}$, and return $\mathcal{V}\left(O(\mathbf{x}_{t})^{'})\right)$. \\
        ~~~~~\If{$\mathcal{V}(O(\mathbf{x}_{t}))$  == false and $\mathcal{V}(O(\mathbf{x}_{t})^{'})$  == true}{
              ~~~~~\Return {Unlearning process has executed.}\\
             }
        ~~~~~\If{$\mathcal{V}(O(\mathbf{x}_{t}))$  == false and $\mathcal{V}(O(\mathbf{x}_{t})^{'})$  == false}{
             ~~~~~\Return {Unlearning process has not executed.}\\
        }   
        \textbf{Model provider executes}: \\
	~~~~~\textbf{Upon} receiving a prediction request for $\mathbf{x}${\\
             ~~~~~~~~~~\Return {$O(\mathbf{x})$}.\\
        }
\end{algorithm}

where, $n$ denotes the total number of training samples, and $p$ denotes the number of samples we intend to modify. For notation simplicity, we assume that the first $p$ training images are modified by introducing a perturbation $\Delta_i$. The objective is to optimize all $\Delta_i$ that the one test sample $\left(\mathbf{x}_t, y_t\right)$ is reclassified with the newly assigned target label $y_{target}$.

Inspired by~\cite{DBLP:conf/iclr/GeipingFHCT0G21}, we can transform the above optimization problem to a gradient matching problem, ensuring that the training gradient aligns with the target gradient:

\begin{equation}
    \begin{aligned}
    \Phi = & \nabla_\theta \ell\left( M\left(\mathbf{x}_t\right),y_{\text {target}}\right) \approx \\ 
    & \frac{1}{p} \sum_{i=1}^p \nabla_\theta \ell\left(M\left(\mathbf{x}_i+\Delta_i\right), y_i\right) =\Psi
    \end{aligned}
\end{equation}

This process ensures that the model's gradient steps, minimizing the training loss on the modified data, $\Psi$, will also minimize the target loss, $\Phi$, which will lead to the misclassification of the test sample $\mathbf{x}_{t}$. Similarly, we also further choose to minimize their negative cosine similarity, since gradient magnitudes vary significantly across different training stages:

\begin{equation}
    \label{equ:generate}
    \mathcal{B}(\Delta, \theta)=1-\frac{\left\langle \Phi, \Psi \right\rangle}{\left\|\Phi\right\| \cdot\left\|\Psi\right\|}
\end{equation}

We do this by optimizing $\Delta$ to achieve the misclassification of one given test sample. Note that if the data provider faces limitations in computational capacity, any trustworthy third party can undertake this task. This configuration is quite common in existing verification schemes~\cite{TIFSYu2024,DBLP:journals/popets/SommerSWM22}.

\subsection{Unlearning Process Verification}
Section~\ref{sec:influential_sample_pair_generation} mainly explains how to generate $(\mathcal{D}_u,\mathbf{x}_t)$ such that the presence of samples $\mathcal{D}_u$ will affect $\mathbf{x}_t$, leading to the misclassification of $\mathbf{x}_t$. In this section, we will describe how to implement the verification of machine unlearning based on generated $(\mathcal{D}_u,\mathbf{x}_t)$.

\subsubsection{Generation and Injection} For the purpose of unlearning verification, the data provider needs to generate the influential sample pairs, and then send the trigger samples along with their own dataset to MLaaS for model training. Then, the data provider sends reaction samples $\mathbf{x_t}$ and asks the model provider what the output of those samples is. The model provider returns the outputs $O(\mathbf{x_t})$ of those samples. After that, the data provider uses a pre-prepared verification scheme to check whether the sample with the output $O(\mathbf{x_t})$ is in the model. The verification scheme will output a result of $\mathcal{V}\left(O(\mathbf{x_t})\right)$. In Algorithm~\ref{algorithm:generation_and_injection}, we provide a detailed process for sample generation and injection. 

In Algorithm~\ref{algorithm:generation_and_injection}, we begin by selecting a set of samples $\mathcal{Q}$. Following that, we modify those samples $\mathcal{Q}$ based on the Equation~\ref{equ:generate}, resulting in the misclassification of another test sample $\mathbf{x}_t$ (lines 3). It is important to note that these optimization processes do not alter the labels of $\mathcal{Q}$; they merely introduce minor perturbations to each sample. This minimizes the impact on the model's performance as much as possible. The data provider then sends the modified samples, along with the original training data, to the server, where the model is trained based on this combined dataset (line 8).

\subsubsection{Sending and Executing Unlearning} To request unlearning, the data provider sends an unlearning request regarding samples $\mathcal{D}_{u}$. The model provider locates and deletes samples $\mathcal{D}_{u}$ from the dataset and executes the unlearning process.

\subsubsection{Re-query and verification}  After sending the unlearning request, the data provider asks again what the output of $\mathbf{x_t}$ is. The model provider returns the outputs $O(\mathbf{x_t})^{'}$. The data provider, once again, uses the verification scheme to check whether the sample under the output $O(\mathbf{x_t})^{'}$ is in the model. Normally, the verification scheme will output a result of $\mathcal{V}\left(O(\mathbf{x_t})^{'}\right)$. Data provider compares the relationship between $\mathcal{V}\left(O(\mathbf{x_t})\right)$ and $\mathcal{V}\left(O(\mathbf{x_t})^{'}\right)$, thereby determining whether the model provider has executed the unlearning operation. In Algorithm~\ref{algorithm:verification}, we provide a detailed process for this process.

In lines 5–6 in Algorithm~\ref{algorithm:generation_and_injection}, the data provider initiates the verification process by querying the output for $\mathbf{x}_t$. Normally, the output results a misclassification result, $\mathcal{V}(O(\mathbf{x}_{t}))$ == false. Subsequently, in line 3 in Algorithm~\ref{algorithm:verification}, the data provider will send an unlearning request, followed by another query for the output related to $\mathbf{x}_t$~(lines 4-5). If $\mathbf{x}_t$ is classified correctly~($\mathcal{V}\left(O(\mathbf{x_t})^{'}\right)$ == True), it confirms that the model provider has indeed performed the unlearning operation~(lines 6-7); otherwise, it suggests that the model provider has not executed the unlearning operation~(lines 8-9). IndirectVerify can be further combined with methods, such as hypothesis testing~\cite{DBLP:journals/popets/SommerSWM22}, to enhance its robustness and verification effectiveness. This will be a focus of our future research.

\section{Theoretical Analysis}
\subsection{Effectiveness Analysis}
\label{sec:influential_sample_pair_selection}
IndirectVerify is mainly based on the idea that the presence or absence of trigger samples affects the classification results of reaction samples. In this section, we would like to provide theoretical analyses to illustrate why IndirectVerify is effective. We formalize the change of classification result of reaction samples as the change of loss: when given a test sample, how can we calculate the effect of each training sample's presence on the change of model's loss regarding this test sample? We will give a closed-form expression for this. Based on this expression, we will illustrate how IndirectVerify minimizes this effect to achieve verification purposes.

Formally, we use $\mathcal{D}=\left\{\left(\mathbf{x}_{1}, y_{1}\right), \left(\mathbf{x}_{2}, y_{2}\right),...,\left(\mathbf{x}_{n}, y_{n}\right) \right\}$ to represent the training dataset. Let $ \ell \left(M\left(\mathbf{x}_i\right), y_i\right)$ be the loss, and define $R(\theta) \stackrel{\text { def }}{=} \frac{1}{n}\sum_{i=1}^{n} \ell \left(M\left(\mathbf{x}_i\right), y_i\right)$ as the empirical risk. Assuming the empirical risk minimizer is given by:
\begin{equation}
    \hat{\theta} \stackrel{\text { def }}{=} \arg \min \frac{1}{n} \sum_{i=1}^n \ell \left(M\left(\mathbf{x}_i\right), y_i\right)
\end{equation}



Considering how the model's loss of a test sample would change if we were to remove samples $\mathcal{D}_u$ from the training dataset, Formally, this change is 

\begin{equation}
    \Delta \ell =  \ell\left(M_{\hat{\theta}_{\mathcal{D}_u \notin \mathcal{D}}}\left(\mathbf{x}_{t}\right), y_{t}\right) - \ell \left(M_{\hat{\theta}}\left(\mathbf{x}_{t}\right), y_{t}\right)
\end{equation}

where 
\begin{equation}
    \hat{\theta}_{\mathcal{D}_u \notin \mathcal{D}} \stackrel{\text { def }}{=} \arg \min _{\mathcal{D}_u \notin \mathcal{D}} \frac{1}{n} \sum_{i=1}^n \ell \left(M\left(\mathbf{x}_i\right), y_i\right)
\end{equation}

Directly calculating the above is also prohibitively slow, since retraining the model each time after removing one training sample and observing how the loss for specific samples changes is not feasible. Inspired by the work of influence function~\cite{DBLP:conf/icml/KohL17}, we can compute the above change if $\mathcal{D}_{u}$ were upweighted by some small $\epsilon$, giving us new parameters

\begin{equation}
    \begin{aligned}
    \hat{\theta}_{\epsilon, \mathcal{D}_{u}}  \stackrel{\text { def }}{=} &\arg \min( R(\theta) + \epsilon L(M\left(\mathcal{D}_{u}\right), y_i))\\
    \stackrel{\text { def }}{=} & \arg \min (\frac{1}{n} \sum_{i=1}^n \ell \left(M\left(\mathbf{x}_i\right), y_i\right) \\
    ~&+ \epsilon \sum_{\mathbf{x}_i \in \mathcal{D}_u} \ell \left(M\left(\mathbf{x}_i\right), y_i\right))\\
    \end{aligned}
\end{equation}

Then the change $\Delta \ell$ will be 

\begin{equation}
    \Delta \ell =  \ell\left(M_{\hat{\theta}_{\epsilon, \mathcal{D}_{u}}}\left(\mathbf{x}_{t}\right), y_{t}\right) - \ell \left(M_{\hat{\theta}}\left(\mathbf{x}_{t}\right), y_{t}\right),
\end{equation}

when we consider that $\mathcal{D}_{u}$ is upweighted by some small $\epsilon$. Note that, as $\hat{\theta}$ doesn't depend on $\epsilon$, $\Delta \ell $ can be expressed as the impact of changes in $\epsilon$ on the loss, that is:

\begin{equation}
    \begin{aligned}
    \Delta \ell &= \left.\frac{\mathrm{d \ell}\left(M_{\hat{\theta}_{\epsilon, \mathcal{D}_{u}}}\left(\mathbf{x}_{t}\right), y_{t}\right)}{\mathrm{d} \epsilon}\right|_{\epsilon=0} \\ &=\left.\frac{\partial \ell \left(M_{\hat{\theta}_{\epsilon, \mathcal{D}_{u}}}\left(\mathbf{x}_{t}\right), y_{t}\right)} {\partial \hat{\theta}_{\epsilon, \mathcal{D}_u}} \frac{\partial \hat{\theta}_{\epsilon, \mathcal{D}_u}}{\partial \epsilon}\right|_{\epsilon=0} \\
    & =\left.\nabla_\theta \ell \left(M_{\hat{\theta}}\left(\mathbf{x}_{t}\right), y_{t}\right)^{\top} \frac{d \hat{\theta}_{\epsilon, \mathcal{D}_u}}{d \epsilon}\right|_{\epsilon=0}
    \end{aligned}
\end{equation}

Since $\hat{\theta}_{\epsilon, \mathcal{D}_{u}}$ is a minimizer of empirical risk, that is:
\begin{equation}
     \nabla R_{\mathcal{D}}\left(\hat{\theta}_{\epsilon, \mathcal{D}_{u}}\right)+\epsilon \nabla L_{\mathcal{D}_u}\left(\hat{\theta}_{\epsilon, \mathcal{D}_{u}}\right) = 0
\end{equation}

Meanwhile, since $\hat{\theta}_{\epsilon, \mathcal{D}_u} \rightarrow \hat{\theta}$ as $\epsilon \rightarrow 0$, we can conduct a Taylor expansion of the left-hand side:
\begin{equation}
    \begin{aligned}
    & {[\nabla R_{\mathcal{D}}(\hat{\theta})+\epsilon \nabla L_{\mathcal{D}_u}(\hat{\theta})]+}  \\
    & {\left[\nabla^2 R_{\mathcal{D}}(\hat{\theta})+\epsilon \nabla^2 L_{\mathcal{D}_u}(\hat{\theta})\right] \Delta_\epsilon} \approx 0 ,
    \end{aligned}
\end{equation}

where we have omitted terms of order $o\left(\left\|\Delta_\epsilon\right\|\right)$.

Finding the solution for $\Delta_\epsilon$ yields:

\begin{equation}
\begin{gathered}
    \Delta_\epsilon \approx- \left[\nabla^2 R_{\mathcal{D}}(\hat{\theta})+\epsilon \nabla^2 L_{\mathcal{D}_u}(\hat{\theta})\right]^{-1} \\
    {[\nabla R_{\mathcal{D}}(\hat{\theta})+\epsilon \nabla L_{\mathcal{D}_u}(\hat{\theta})] .}
\end{gathered}
\end{equation}

As $\hat{\theta}$ minimizes $R$, we have $\nabla R(\hat{\theta})=0$. Neglecting terms of order $o(\epsilon)$, we obtain:

\begin{equation}
    \Delta_\epsilon \approx-\nabla^2 R(\hat{\theta})^{-1} \nabla L_{\mathcal{D}_u}(\mathbf{x}_{i},\hat{\theta}) \epsilon .
\end{equation}

We further make the assumption that the empirical risk is both twice-differentiable and strongly convex, and the following

\begin{equation}
    H_{\hat{\theta}} \stackrel{\text { def }}{=} \nabla^2 R(\hat{\theta})=\frac{1}{n} \sum_{i=1}^n \nabla_\theta^2 \ell \left(M\left(\mathbf{x}_{i}\right), y_{i}\right)
\end{equation}

exists and is positive-definite. Therefore, we conclude that:

\begin{equation}
    \left.\frac{d \hat{\theta}_{\epsilon, z}}{d \epsilon}\right|_{\epsilon=0} =-H_{\hat{\theta}}^{-1} \nabla L_{\mathcal{D}_u}(\mathbf{x}_{i},\hat{\theta})
\end{equation}

Given that removing a sample is equivalent to upweighting it by $\epsilon=-\frac{1}{n}$, we can linearly estimate the change in loss resulting from the removal of $\mathcal{D}_u$ without retraining the model by computing:

\begin{equation}
    \label{equation:final}
    \Delta \ell  \approx-\frac{1}{n} \nabla_\theta \ell \left(M_{\hat{\theta}}\left(\mathbf{x}_{t}\right), y_{t}\right)^{\top} H_{\hat{\theta}}^{-1} \nabla L_{\mathcal{D}_u}(\mathbf{x}_{i},\hat{\theta}).
\end{equation}

$\Delta \ell$ describes how, when a test sample $\mathbf{x}_{t}$ is provided, removing a training sample $\mathbf{x}_{i}$ influences the loss of $\mathbf{x}_{t}$. Consider any training dataset, with no external manipulation involved; if we want to find influential sample pairs, we can first identify misclassified test samples, satisfying the criteria:

\begin{equation}
    \label{equation:final_search_1}
    \begin{aligned}
    M_{\hat{\theta}}(\mathbf{x}_{t}) & = \hat{y_{t}} \neq y_{t}, \\
    \text{s.t.}  \quad &\hat{\theta} = \arg \min_{\mathcal{D}_{u} \in \mathcal{D}}  \frac{1}{n} \sum_{i=1}^n \ell \left(M\left(\mathbf{x}_i\right), y_i\right) \\
    \end{aligned}
\end{equation}

Subsequently, we can index the training dataset to find the sample whose removal minimizes the loss of that test sample, that is, $\min(\Delta \ell)$, and thereby:
\begin{equation}
    \label{equation:final_search_2}
    \begin{aligned}
     M_{\hat{\theta}_{\mathcal{D}_u \notin \mathcal{D}}}& (\mathbf{x}_{t}) = \hat{y_{t}} = y_{t}, \\
    \text{s.t.}  \quad &\hat{\theta}_{\mathcal{D}_u \notin \mathcal{D}} = \arg \min_{\mathcal{D}_{u} \notin \mathcal{D}} \frac{1}{n} \sum_{i=1}^n \ell \left(M\left(\mathbf{x}_i\right), y_i\right) \\
    \end{aligned}
\end{equation}

However, directly finding the influential sample pairs based on the above equation is very difficult, especially when faced with complex models and enormous training data~(see the experimental results in Figure~\ref{fig:influence_cifar_harmful}).

To ensure the persistent existence of such influential sample pairs, we would like to directly modify specific samples to $\min(\Delta \ell)$. That is, for a given reaction sample, we want to modify a set of trigger samples such that removing those trigger samples will:

\begin{equation}
    \begin{aligned}
     & \min\left(\ell\left(M_{\hat{\theta}_{\mathcal{D}_u \notin \mathcal{D}}}\left(\mathbf{x}_{t}\right), y_{t}\right) - \ell \left(M_{\hat{\theta}}\left(\mathbf{x}_{t}\right), y_{t}\right)\right)\\
     & \text{s.t.} M_{\hat{\theta}_{\mathcal{D}_u \notin \mathcal{D}}}(\mathbf{x}_{t}) == y_{t}, M_{\hat{\theta}}(\mathbf{x}_{t}) \neq y_{t}
    \end{aligned}
\end{equation}

In general, minimizing $\Delta \ell$ can contain two methods: maximizing $\ell \left(M_{\hat{\theta}}\left(\mathbf{x}_{t}\right), y_{t}\right)$
or minimizing $\ell\left(M_{\hat{\theta}_{\mathcal{D}_u \notin \mathcal{D}}}\left(\mathbf{x}_{t}\right), y_{t}\right)$. Therefore, we first choose a test sample that is classified correctly, which ensures minimizing $\ell\left(M_{\hat{\theta}_{\mathcal{D}_u \notin \mathcal{D}}}\left(\mathbf{x}_{t}\right), y_{t}\right)$. Then, perturb some training samples, which will maximize $\ell \left(M_{\hat{\theta}}\left(\mathbf{x}_{t}\right), y_{t}\right)$. This perturbation is designed to induce a high loss for the associated test sample and result in misclassification. Subsequently, upon removing these modified samples from the model, we will achieve $\min(\Delta \ell)$, facilitating the recovery of the original labels for the given test sample.

\begin{figure*}
    \centering
    \includegraphics[width=0.75\linewidth]{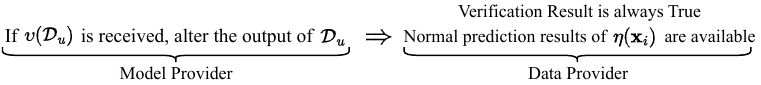}
    \caption{The Illustration of Determinism for Existing Verification Schemes.}
    \label{fig:robustness1}
\end{figure*}

\begin{figure*}
    \centering
    \includegraphics[width=0.75\linewidth]{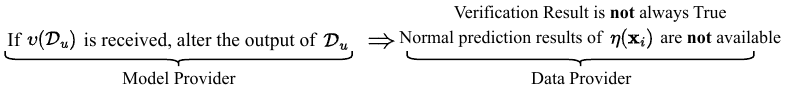}
    \caption{The Illustration of Determinism for Our Verification Schemes.}
    \label{fig:robustness2}
\end{figure*}

\subsection{Robustness Analysis}
Our analysis mainly focuses on how to ensure the determinism of the output of the test sample $\mathbf{x}_t$ used for verification, as well as the confidentiality of the relationship between $\mathcal{D}_u$ and $\mathbf{x}_t$. We respectively use $\eta(\cdot)$ and $\upsilon(\cdot)$ to represent prediction and unlearning requests.

\begin{itemize}

    \item \textbf{The Determinism of the Output of the Verification Test Sample $\mathbf{x}_t$ is Achieved.} Existing verification schemes, such as~\cite{DBLP:conf/aaai/GravesNG21,DBLP:conf/iwqos/LiuMYWL21,DBLP:conf/eccv/GolatkarAS20,DBLP:conf/cvpr/GolatkarARPS21,TIFSYu2024,DBLP:journals/popets/SommerSWM22}, typically involve sending $\upsilon(\mathcal{D}_u)$ and then observing $\eta(\mathcal{D}_u)$.  In this situation, as shown in Figure~\ref{fig:robustness1}, the model provider can introduce various fine-tuning strategies to ensure that the fine-tuned model’s outputs bypass the verification process without any effect on other query requests.


    IndirectVerify involves sending $\upsilon(\mathcal{D}_u)$ and then observing $\eta(\mathbf{x}_t)$. In this case, model provider cannot directly alter the output of sample $\mathcal{D}_u$ in any received unlearning request to bypass the verification. As shown in Figure~\ref{fig:robustness2}, any arbitrary adjustment may impact the model's performance during the querying process, resulting in the misclassification of samples that were originally classified incorrectly. In addition, changing the output of $\mathcal{D}_u$ does not necessarily alter the output of $\mathbf{x}_t$. Therefore, for the verification scheme itself, it remains effective.
    

    Therefore, the model provider must indeed execute the unlearning process to meet the verification request. Any attempt to bypass this process could result in inaccuracies in the model's precision, affecting its overall usability.
    
    \item \textbf{The Confidentiality of the Relationship between $\mathcal{D}_u$ and $\mathbf{x}_t$ is Achieved.} IndirectVerify is to generate corresponding $\mathcal{D}_u$ based on the selected sample $\mathbf{x}_t$. Model providers may aim to find the corresponding reaction samples $\mathbf{x}_t$ by analyzing the $\mathcal{D}_u$. This allows the model provider only to alter the output of the corresponding verification sample $\mathbf{x}_t$, thereby bypassing the verification without affecting other normal queries. However, for the model provider, it is challenging to find $\mathbf{x}_t$ based on the corresponding $\mathcal{D}_u$.

     We use $\varsigma(\cdot)$ to denote the process of generating $\mathcal{D}_u$ through $\mathbf{x}_t$. For any given $\mathbf{x}_t$, generating different $\mathcal{D}_u$ based on the function $\varsigma(\cdot)$ is straightforward. The mapping from $\mathbf{x}_t$ to $\mathcal{D}_u$ is \textit{injective}, as different $\mathbf{x}_t$ values correspond to different $\mathcal{D}_u$. That is, 
     \begin{equation*}
        \text{if } \mathbf{x}_{t,1} \neq \mathbf{x}_{t,2}, \text{ then } \varsigma(\mathbf{x}_{t,1}) \neq \varsigma(\mathbf{x}_{t,2}). 
    \end{equation*}
     However, this mapping is not \textit{surjective} because not all $\mathcal{D}_u \in \mathcal{D}$ have corresponding $\mathbf{x}_t$. That is, 
     \begin{equation*}
        \exists \mathcal{D}_u \in \mathcal{D}, \text{for which}, \text{for any } \mathbf{x}_{t},\varsigma(\mathbf{x}_{t}) \neq \mathcal{D}_u \text{ holds True}.
    \end{equation*}
    Therefore, for model provider, since $\varsigma(\cdot)$ is injective and not surjective, this prevents the model provider from deducing  $\mathbf{x}_t$ based on $\mathcal{D}_u$, thereby ensuring that the model provider cannot bypass the verification process.

\end{itemize}

\section{Performance Evaluation}
\subsection{Experiment Setup}
\subsubsection{Model and Dataset.} IndirectVerify can be applied to all types of classification models. To show its effectiveness, we choose two powerful and popular machine learning models: VGG and ResNet. We adopt five real-world image datasets for evaluation: MNIST~\footnote{http://yann.lecun.com/exdb/mnist/}, Fashion MNIST~\footnote{http://fashion-mnist.s3-website.eu-central-1.amazonaws.com/}, SVHN~\footnote{http://ufldl.stanford.edu/housenumbers/}, CIFAR-10 and CIFAR-100~\footnote{https://www.cs.toronto.edu/~kriz/cifar.html}. The datasets cover different attributes, dimensions, and numbers of categories, allowing us to explore the verification results of the proposed scheme effectively.

\subsubsection{Baseline Methods.} To demonstrate the effectiveness of IndirectVerify, we consider the following baseline schemes:
\begin{itemize}
    \item Backdoor-based verification: Backdoor-based verification scheme has been proposed for verifying the unlearning process in MLaaS~\cite{TIFSYu2024,DBLP:journals/popets/SommerSWM22}. These schemes usually involve training models in advance using samples with embedded backdoors, $\mathcal{D}_{u}$ and subsequently using samples with the same backdoor pattern to detect whether $\mathcal{D}_{u}$ have been successfully unlearned.
    \item MIAs-Based verification: MIAs have already been widely used as one of the methods to measure the effectiveness of machine unlearning scheme~\cite{DBLP:conf/aaai/GravesNG21,DBLP:conf/iwqos/LiuMYWL21,DBLP:conf/eccv/GolatkarAS20,DBLP:conf/cvpr/GolatkarARPS21}. Here, we also include MIAs as one of the baseline methods
    \item Random removing sample: Our verification scheme is based on generating $(\mathcal{D}_u,\mathbf{x}_t)$, where after unlearning $\mathcal{D}_u$ from the model, the model's loss for $\mathbf{x}_t$ will decrease. To compare, in Section~\ref{sec:feasibility}, we consider the impact on $\mathbf{x}_t$ after unlearning randomly selected $\mathcal{D}_u$ from the model.
\end{itemize}

\begin{figure*}[!t]
      \centering
      \subfloat[ResNet CIFAR-10]{\includegraphics[width=0.25\linewidth]{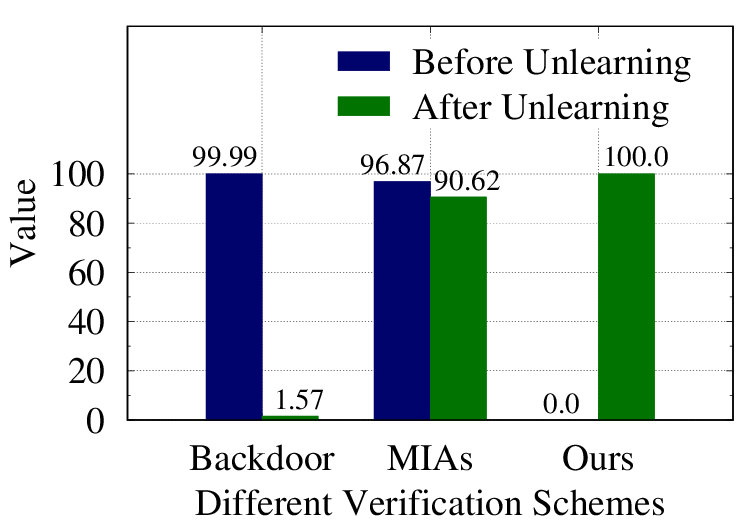}\label{fig:vrresnet_cifar10}}
      \subfloat[ResNet CIFAR-100]{\includegraphics[width=0.25\linewidth]{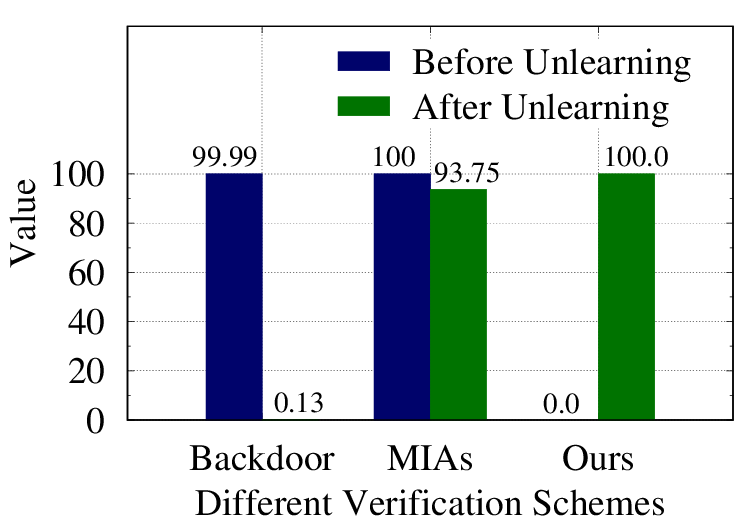}\label{fig:vrresnet_cifar100}}
      \subfloat[ResNet SVHN]{\includegraphics[width=0.25\linewidth]{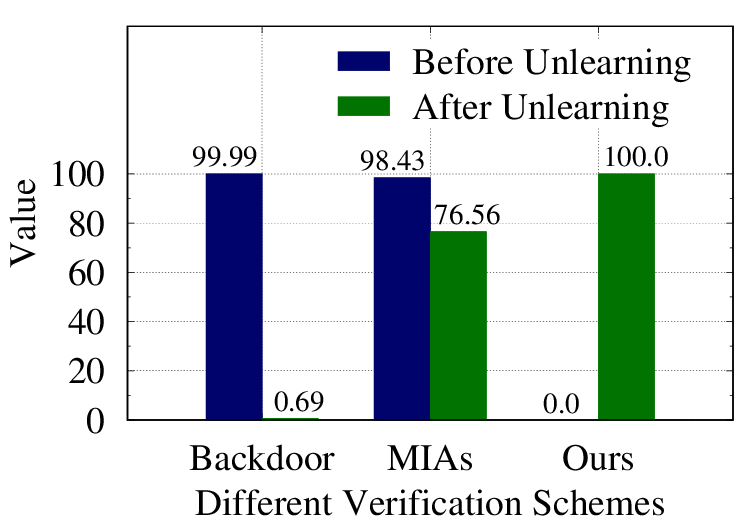}\label{fig:vrresnet_svhn}}
      \subfloat[VGG CIFAR-10]{\includegraphics[width=0.25\linewidth]{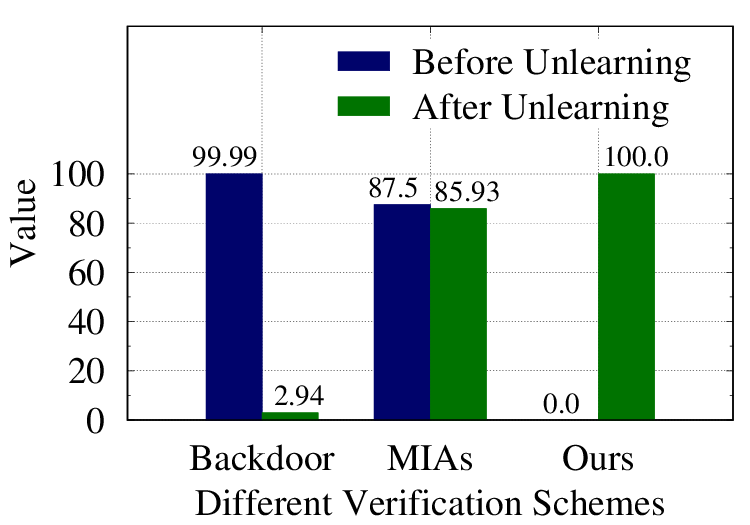}\label{fig:vrvgg_cifar10}}\\
      \subfloat[VGG CIFAR-100]{\includegraphics[width=0.25\linewidth]{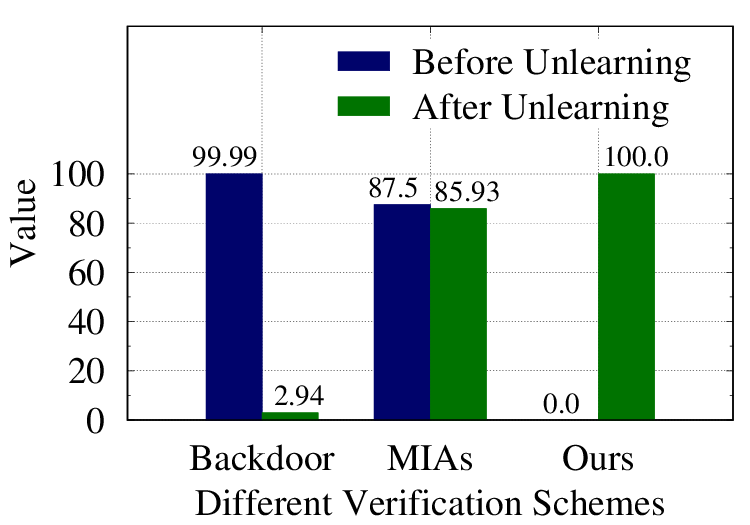}\label{fig:vrvgg_cifar100}}
      \subfloat[VGG SVHN]{\includegraphics[width=0.25\linewidth]{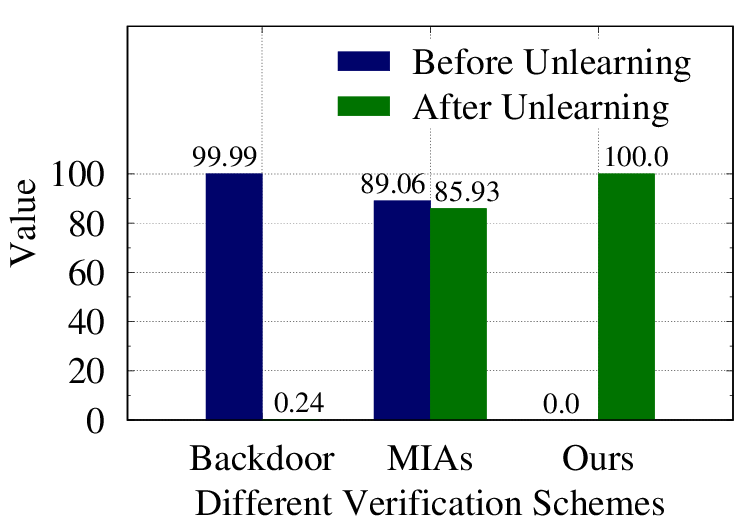}\label{fig:vrvgg_svhn}}
      \subfloat[Fashion MNIST]{\includegraphics[width=0.25\linewidth]{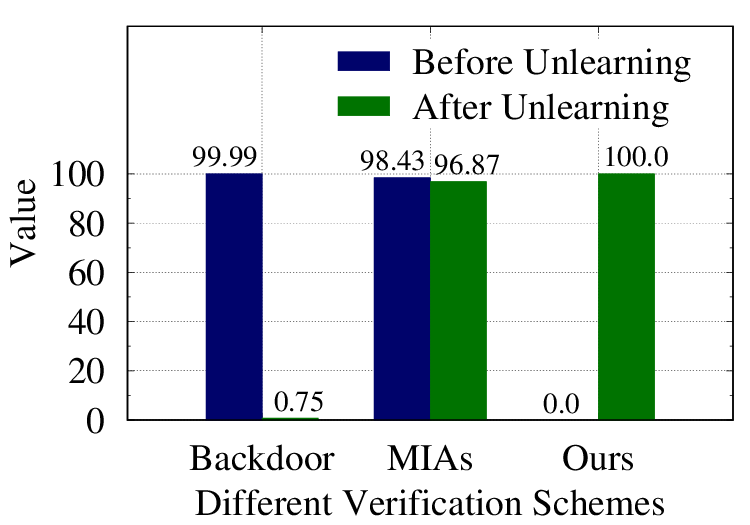}\label{fig:vrfmnist}}
      \subfloat[MNIST]{\includegraphics[width=0.25\linewidth]{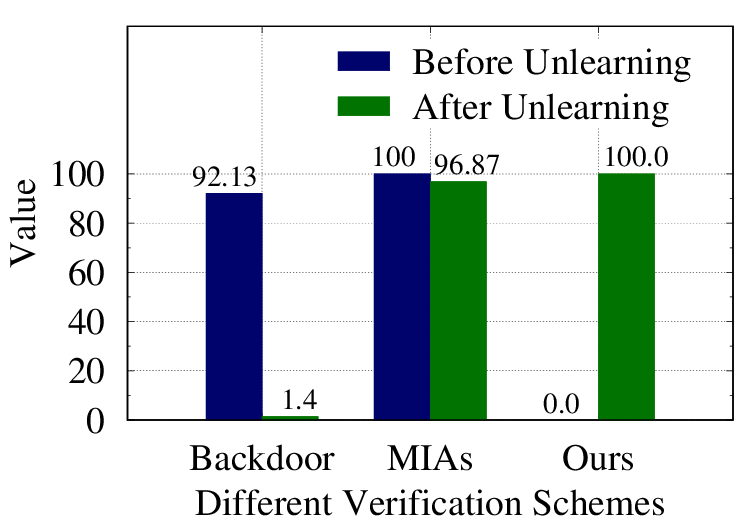}\label{fig:vrmnist}}\\
    \caption{Verification results of different schemes. The Y-axis represents \textit{ACC}, \textit{INA} and \textit{accuracy} on the selected test sample, respectively. In the backdoor scheme, embedding initially yields high accuracy, reduced to near 0\% after unlearning, confirming successful removal. MIAs-based verification effectively identifies training samples but fails post-unlearning. IndirectVerify, using influential sample pairs, can verify the unlearning process effectively.}%
    \label{fig:Verifiactionresults}
\end{figure*} 

\subsubsection{Metrics} 
For backdoor-based verification schemes, we measure whether the model correctly classifies samples with the backdoor pattern as the target label, that is, $ACC = \frac{TP}{TP + FN}$, where $TP$ denotes the number of backdoored samples predicted as the target label and $TP + FN$ represents the total test backdoored samples. Ideally, before unlearning, $ACC$ should be close to $100\%$. And after unlearning, $ACC$ should approach $0\%$. For MIAs-based verification schemes, we use the success rate of attacking unlearning samples as our metric, that is, $INA = \frac{TP}{TP + FN}$, where $TP$ denotes the number of unlearning samples predicted to be in the training set and $TP + FN$ represents the total unlearning samples. Ideally, $INA$ on these unlearning samples should be close to 100\% before unlearning. After unlearning, $INA$ should be close to $0\%$. For our scheme, we evaluate based on whether the test samples are classified correctly. Ideally, before unlearning, the classification accuracy of the test sample should be close to $0\%$. After unlearning, the accuracy should be close to $100\%$.

\subsection{Verification Results}
\label{sec:verificationresults}

\textbf{Setup.} To evaluate IndirectVerify, we will thoroughly compare IndirectVerify with backdoor-based and MIAs-based verification schemes. We adopt retraining from scratch as our unlearning scheme, aligning with experiments conducted in existing works~\cite{TIFSYu2024,DBLP:journals/popets/SommerSWM22}.
\begin{itemize}
    \item For the backdoor-based verification scheme~\cite{TIFSYu2024,DBLP:journals/popets/SommerSWM22}, we use the same method described in~\cite{TIFSYu2024} and insert backdoor into $10\%$ of the samples during the training process. Following this, we execute the unlearning process and record the $ACC$ before and after the unlearning process. 
    \item For the MIAs-based verification scheme, we do our evaluation following the approach outlined in~\cite{DBLP:conf/uss/LiuWH000CF022}. In the case of the CIFAR100 dataset, we choose the top $10$ outputs as inputs for the attack model. We record the $INA$ before and after the unlearning process. 
    \item For our scheme, we begin by choosing one specific test sample. Subsequently, we perturb $5\%$ of the training dataset using Equation~\ref{equ:generate}. We then measure the model's classification accuracy on the selected test sample, both with and without these perturbed training samples. 
\end{itemize} 

The results are shown in Figure~\ref{fig:Verifiactionresults}. Each sub-figure illustrates the results of various verification schemes applied to various models and datasets. 

\textbf{Results.} For the backdoor-based scheme, after embedding the backdoor, the accuracy of correctly classifying samples with the backdoor as the target is very high. This indicates that the model is indeed trained using the backdoor samples. After performing the unlearning operation, the classification results of samples with the backdoor are close to $0\%$. This suggests that the backdoor-based verification scheme successfully confirms that the model has indeed undergone an unlearning process, resulting in the removal of backdoor samples from the model and causing a decrease in the accuracy of samples with the backdoor. For MIAs-Based verification schemes, before the unlearning process, all success rates are relatively high, indicating that MIAs successfully identify samples used for model training. However, after unlearning, the attack accuracy does not show a significant decrease. This illustrates that MIAs cannot identify samples that are not in the training set. This result suggests that, given the limited effectiveness, MIAs may not be suitable for verifying the unlearning process~\cite{DBLP:journals/csur/HuSSDYZ22}.

\begin{figure*}[!t]
      \centering
      \subfloat[Removing Harmful Samples~(MNIST)]{\includegraphics[width=0.33\linewidth]{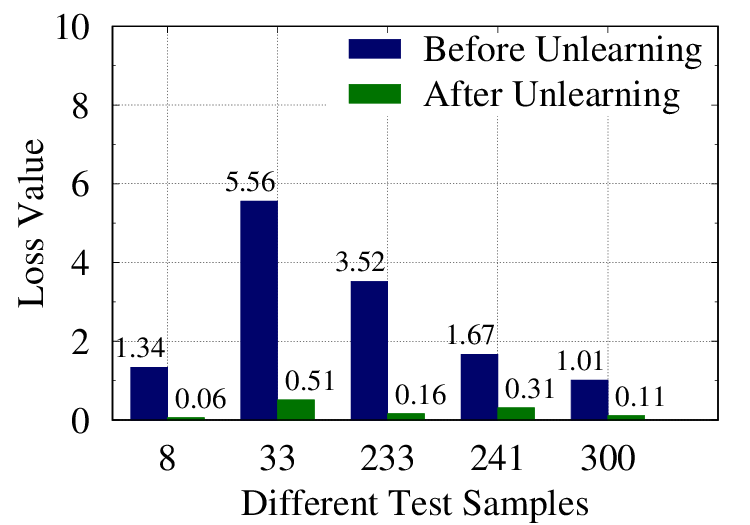}\label{fig:influence_mnist_harmful}}
      \subfloat[Random Removing Samples(MNIST)]{\includegraphics[width=0.33\linewidth]{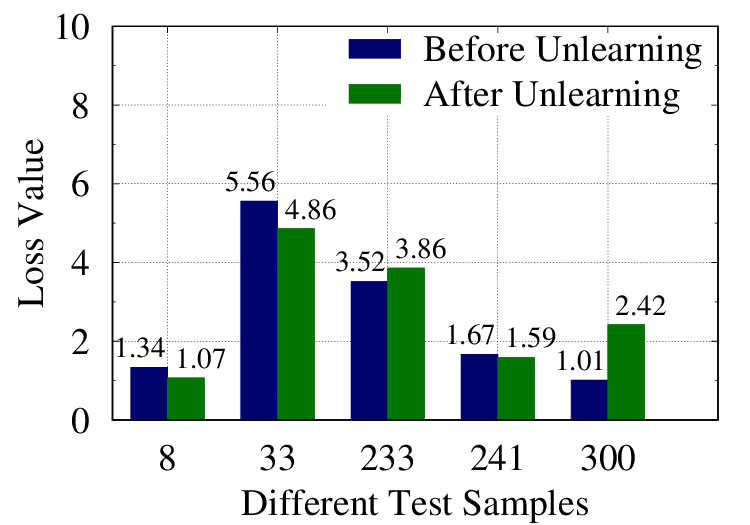}\label{fig:influence_random}}
      \subfloat[Removing Harmful Samples~(CIFAR10)]{\includegraphics[width=0.33\linewidth]{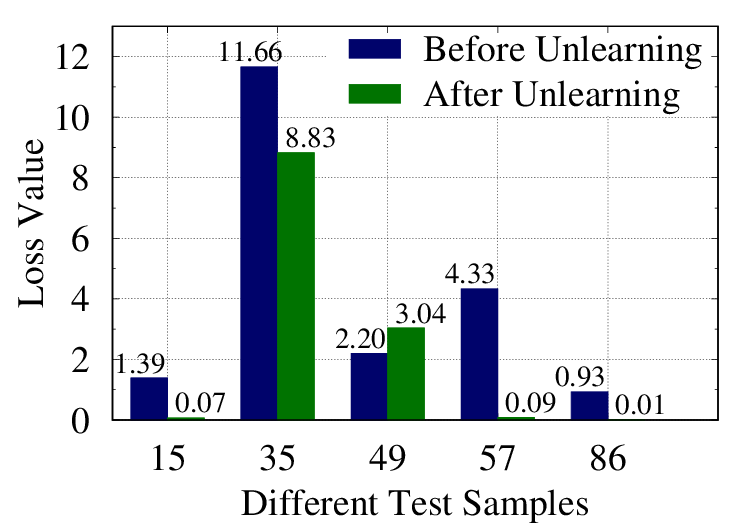}\label{fig:influence_cifar_harmful}}
      \caption{The loss of each sample before and after removing different types of samples. The x-axis denotes various test samples, while the y-axis represents the corresponding loss. For MNIST, removing the most harmful samples significantly reduces loss, while random removal has no effect. For CIFAR-10, removing selected harmful training samples doesn't decrease test sample loss. Possible reasons include a lack of abnormal samples or the ineffectiveness of Equation~\ref{equation:final} assumptions in complex models. To address the consistent presence of harmful samples in the dataset, we propose a pre-embedding method in Section~\ref{sec:influential_sample_pair_generation}. It can be applied to more complex models and dataset scenarios.}
      \label{fig:influencefunctionresults}
\end{figure*} 

\begin{figure}
    \centering
    \includegraphics[width=0.85\linewidth]{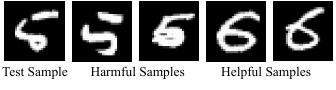}
    \caption{Harmful and helpful samples. we present a misclassified test sample (labeled 5 but classified as 6) alongside its harmful samples (all labeled 5) and helpful samples (all labeled 6). The harmful samples resemble 6 more than 5, contributing to the misclassification of the test sample. Removing these harmful samples eliminates their impact, allowing the helpful samples to better differentiate the test sample and result in the correct classification.}
    \label{fig:influencesamples}
\end{figure}

For our scheme IndirectVerify, before unlearning, all test samples are misclassified due to trigger samples in the training set. At the same time, upon unlearning these trigger samples from the model, the test samples are all classified correctly. This indicates that IndirectVerify is effective for verifying the unlearning. It is noteworthy that our trigger samples are different from backdoored samples. Firstly, we did not alter the labels of those trigger samples, effectively minimizing the impact on the model performance. Secondly, in practical scenarios, model providers do not allow data providers to add backdoors, as it poses a serious threat to the model's security. IndirectVerify, being without backdoors, allows for verification of the unlearning operation while ensuring model security.

\subsection{Feasibility Analysis}
\label{sec:feasibility}

\begin{figure*}[!t]
      \centering
      \subfloat[Outputs before Fine-tuing (MIAs)]{\includegraphics[width=0.25\linewidth]{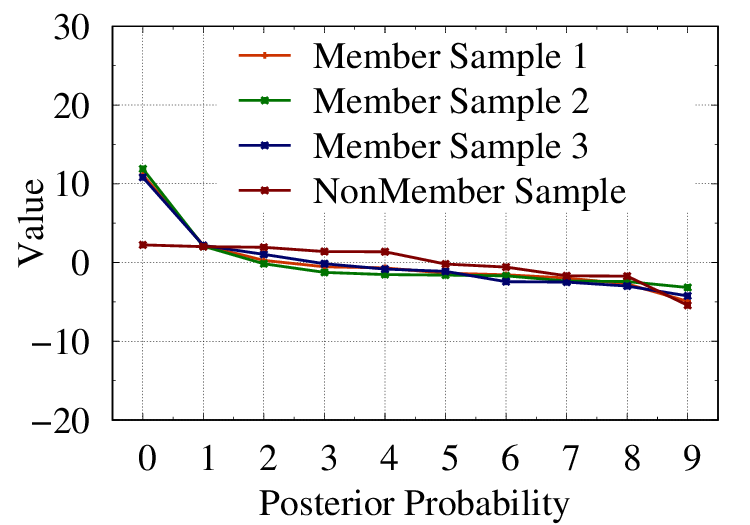}\label{fig:attackmia_outputs_before}}
      \subfloat[Outputs after Fine-tuing (MIAs)]{\includegraphics[width=0.25\linewidth]{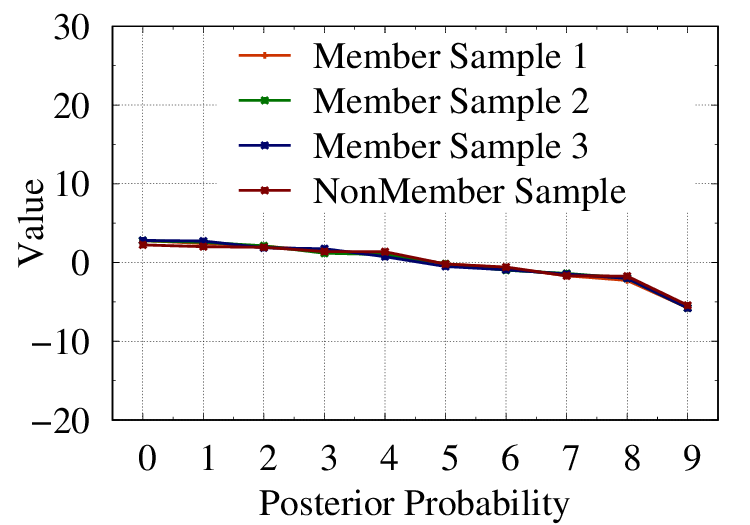}\label{fig:attackmia_outputs_after}}
      \subfloat[Bypassing Results for MIAs]{\includegraphics[width=0.25\linewidth]{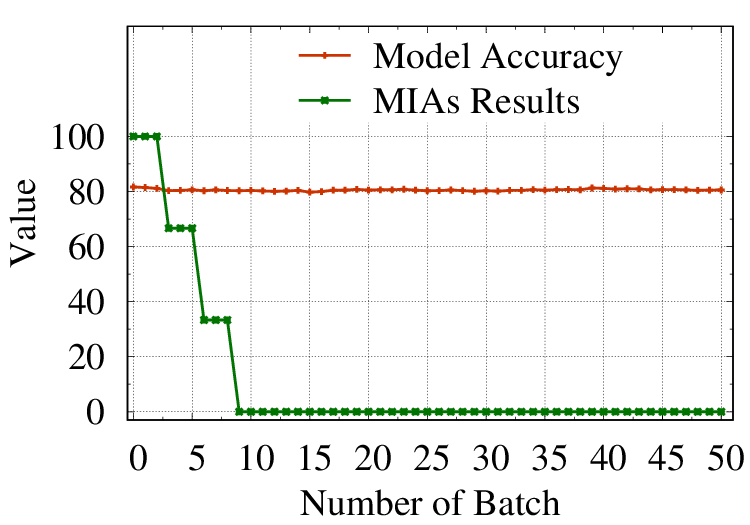}\label{fig:attackmia_results}}
      \subfloat[Bypassing Results for Backdoor]{\includegraphics[width=0.25\linewidth]{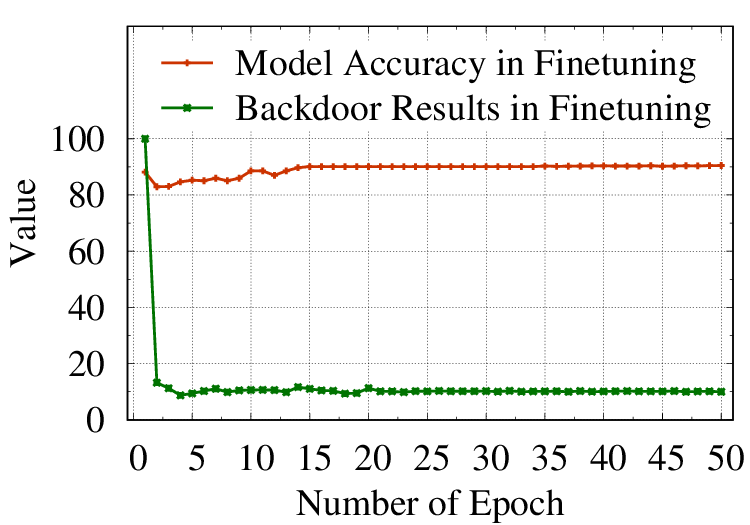}\label{fig:backdoor_results}}
    \caption{The results of our proposed bypassing schemes for MIAs-based and backdoor-based verification schemes. In Figure~\ref{fig:attackmia_outputs_before}, before fine-tuning, the model's outputs for member and non-member samples differ, enabling MIAs to verify the unlearning process based on the different outputs. However, in Figure~\ref{fig:attackmia_outputs_after}, the bypass method based on fine-tuning makes the outputs of member and non-member samples indistinguishable, rendering MIAs ineffective~(see Figure~\ref{fig:attackmia_results}). For backdoor-based verification schemes (Figure~\ref{fig:backdoor_results}), post-training, the model effectively classifies samples with a backdoor. After fine-tuning, the model fails to correctly classify samples with backdoors, making the backdoor-based verification scheme ineffective.}%
    \label{fig:Bypassing_Schemes_for_others}
\end{figure*} 

\begin{figure*}[!t]
      \centering
      \subfloat[Outputs before Fine-tuing]{\includegraphics[width=0.25\linewidth]{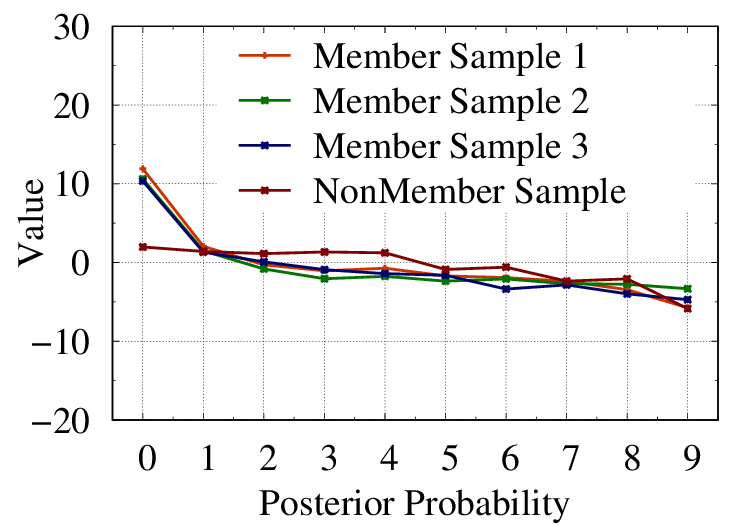}\label{fig:attackour_outputs_before}}
      \subfloat[Outputs after Fine-tuing]{\includegraphics[width=0.25\linewidth]{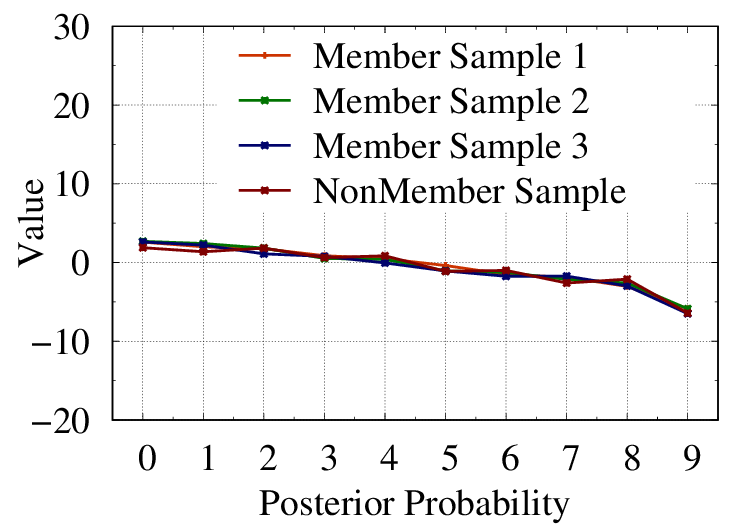}\label{fig:attackour_outputs_after}}
      \subfloat[\centering Bypassing Results (Same setting with MVS)]{\includegraphics[width=0.25\linewidth]{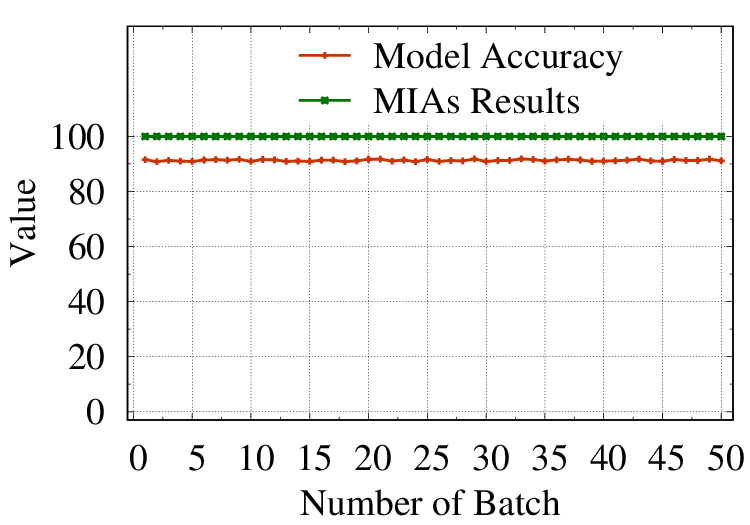}\label{fig:ourmia_results}}
      \subfloat[\centering Bypassing Results (Same setting with BVS)]{\includegraphics[width=0.25\linewidth]{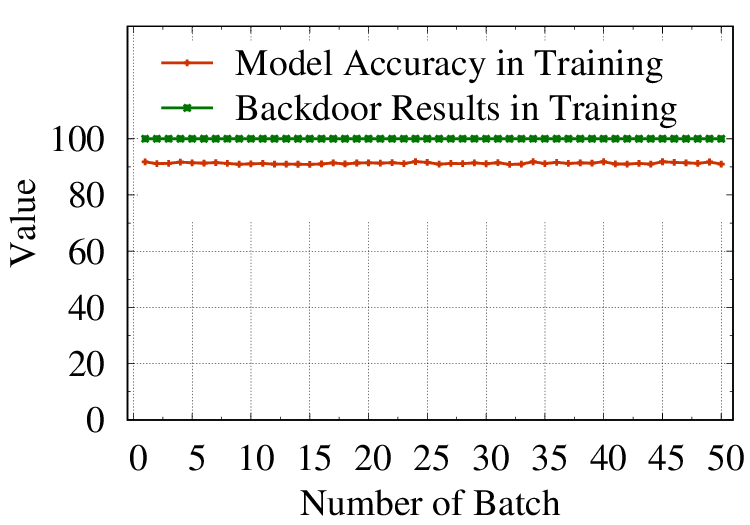}\label{fig:ourbackdoor_results}}
      \caption{The results of our proposed bypassing schemes for our verification scheme. MVS represents the MIAs-based verification scheme, and BVS denotes the backdoor-based verification scheme. Despite closely aligning with one test sample's output after fine-tuning (Figures~\ref{fig:attackour_outputs_before} and ~\ref{fig:attackour_outputs_after}), our verification results (Figure~\ref{fig:ourmia_results}) show no change, indicating effective resistance against fine-tuning bypass schemes based on Equation~\ref{equation:attack_loss}. Similarly, IndirectVerify also remains effective against bypass schemes based on Equation~\ref{equation:attack_loss_for_backdoor_random} (Figure~\ref{fig:ourbackdoor_results}). Overall, IndirectVerify robustly resists known bypass schemes.}%
      \label{fig:Bypassing_Schemes_for_our}
\end{figure*}

\textbf{Setup.} IndirectVerify aims at generating a pair of samples $(\mathcal{D}_u,\mathbf{x}_t)$. Samples $\mathcal{D}_u$ can influence the accuracy of another sample $\mathbf{x}_t$. Removing $\mathcal{D}_u$ corrects the misclassification of sample $\mathbf{x}_t$. In this section, we validate the above points by directly searching samples $\mathcal{D}_u$ based on Equation~\ref{equation:final_search_1} and~\ref{equation:final_search_2} within the dataset, rather than generating them. We first construct one simple model and train it based on the MNIST dataset. Then, we select five misclassified samples and calculate the impact of each training sample on each test sample based on Equation~\ref{equation:final}. We sort the values of the results of Equation~\ref{equation:final} in ascending order, and then select the corresponding training samples. If the value is significant, we refer to the corresponding sample as \textit{helpful}; conversely, if the value is small, we label the corresponding sample as \textit{harmful}. Subsequently, we remove the topmost harmful sample corresponding to each misclassified test sample from the dataset and execute the unlearning process. Once each model is retrained, we retest these misclassified samples using the corresponding unlearned model. As a comparison, we also retrain the model when randomly removing one training sample. In addition, we choose the ResNet18 architecture for the CIFAR10 dataset to do the same process as above. We record the loss of different test samples before and after the unlearning process; the results are shown in Figure~\ref{fig:influencefunctionresults}.

\textbf{Results.} In Figure~\ref{fig:influencefunctionresults}, the x-axis represents different test samples, while the y-axis denotes the corresponding loss of each test sample. Figures~\ref{fig:influence_mnist_harmful} and~\ref{fig:influence_cifar_harmful} respectively illustrate the results of removing the most harmful samples from the MNIST and CIFAR10 datasets. Figure~\ref{fig:influence_random} shows the results when randomly removing one sample from the MNIST dataset. As shown in Figure~\ref{fig:influence_mnist_harmful}, the loss for each misclassified test sample is relatively high before retraining the model. However, upon removing the selected most harmful sample, the test sample's loss significantly decreases, enabling correct classification. Meanwhile, as shown in Figure~\ref{fig:influence_random}, when we randomly remove a sample, the unlearned model does not show a decrease in the loss of each test sample. The above results demonstrate that when dealing with a simple model and dataset, given a misclassified test sample, we can identify the most harmful training sample for this misclassified sample. Removing the harmful training sample will reduce the loss for the corresponding misclassified test sample, leading to a correct classification. Those results illustrate the feasibility of our verification scheme.

In Figure~\ref{fig:influencesamples}, we show a misclassified test sample (labeled 5, but classified as $6$) along with its corresponding harmful samples~(all labeled $5$) and helpful samples~(all labeled $6$). These harmful samples appear dissimilar to $5$ but are closely similar to $6$. The misclassified sample shares significant similarities with these harmful samples. In contrast, the most helpful samples exhibit true characteristics of $6$, aiming to differentiate the misclassified sample as much as possible. The reason for the misclassification of the test sample may be attributed to these harmful training samples, which causes the test sample to be closer to the data space of $6$, resulting in the misclassification. When we remove those harmful samples, the impact of these harmful samples is eliminated, and those helpful samples strive to differentiate the test sample as much as possible, leading to the correct classification of the original test sample.

As shown in Figure~\ref{fig:influence_cifar_harmful}, when removing the selected harmful training samples for CIFAR-10, the loss of the test sample does not decrease. We think there are several possible reasons for this. On one hand, the dataset may lack abnormal samples, resulting in the absence of harmful samples. On the other hand, when we derive Equation~\ref{equation:final}, we make the assumption that the empirical risk is both twice-differentiable and strongly convex. But this may not hold in complex models with large datasets, making Equation~\ref{equation:final} ineffective. Therefore, in Section~\ref{sec:influential_sample_pair_generation}, we have proposed an optimization-based method that is applicable to any complex model and dataset.

\begin{figure*}[!t]
      \centering
      \subfloat[ResNet CIFAR-10]{\includegraphics[width=0.25\linewidth]{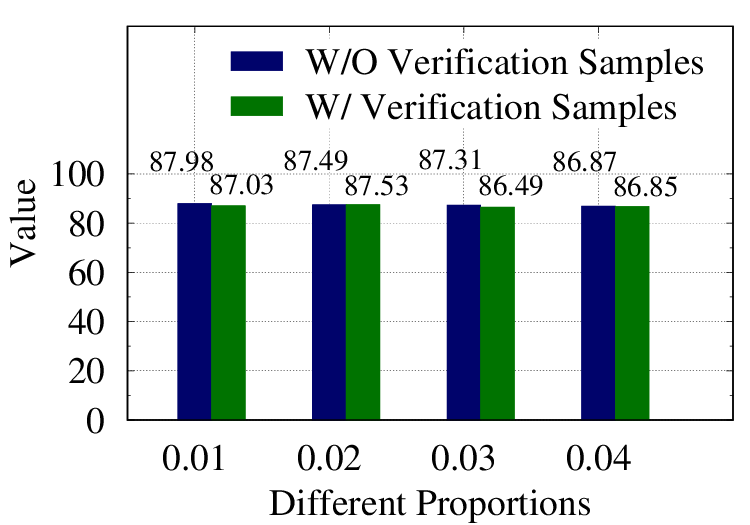}\label{fig:modelaccuracy_cifar10_resnet}}
      \subfloat[VGG CIFAR-10]{\includegraphics[width=0.25\linewidth]{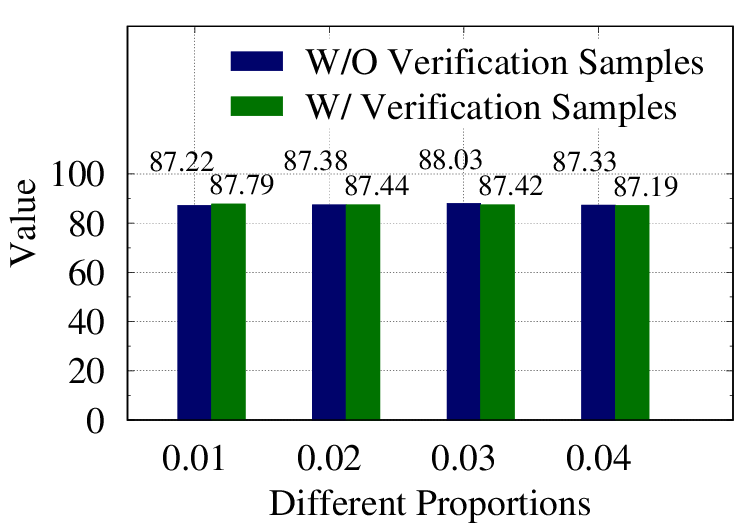}\label{fig:modelaccuracy_cifar10_vgg}}
      \subfloat[ResNet SVHN]{\includegraphics[width=0.25\linewidth]{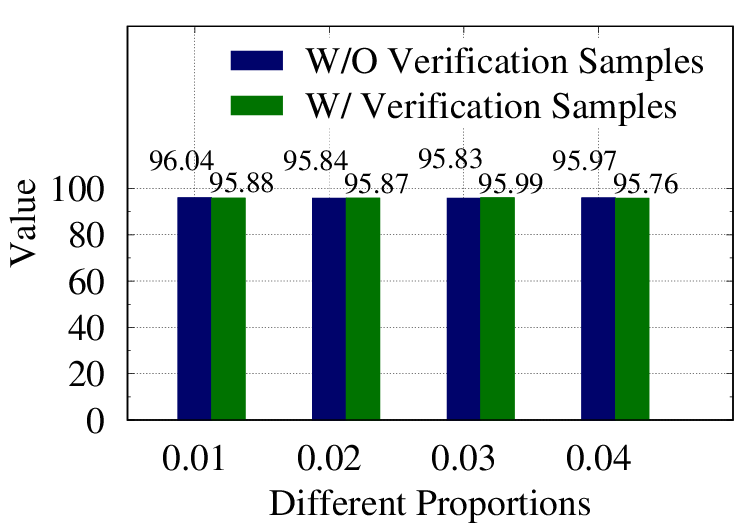}\label{fig:modelaccuracy_svhn_resnet}}
      \subfloat[VGG SVHN]{\includegraphics[width=0.25\linewidth]{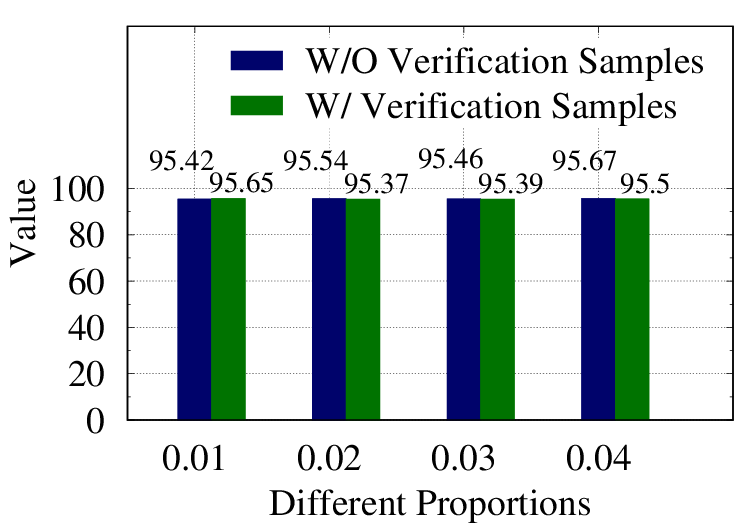}\label{fig:modelaccuracy_svhn_vgg}}\\
      \caption{Model Accuracy with different perturbed trigger samples. All figures show that embedding trigger samples has a negligible impact on model performance. For instance, with the ResNet model and CIFAR-10 dataset, the difference before and after embedding is less than $0.01\%$, likely due to training process randomness. The minimal effect stems from modifying pixel values without changing samples' labels.}%
    \label{fig:modelaccuracy}
\end{figure*} 

\begin{figure}[!t]
      \centering
      \subfloat[With Modification]{\includegraphics[width=0.49\linewidth]{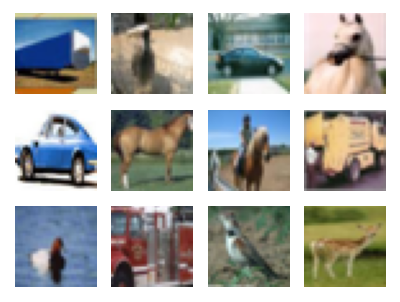}\label{fig:verificationimage_before}}
      \subfloat[Without Modification]{\includegraphics[width=0.49\linewidth]{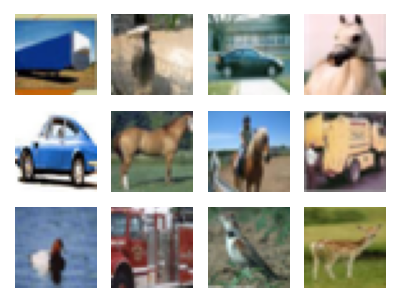}\label{fig:verificationimage_after}}
    \caption{Original samples and the corresponding trigger samples. It is noted that the samples, before and after perturbation, are almost the same, with no significant difference.}%
    \label{fig:verificationimage}
\end{figure}

\subsection{Bypassing}
\label{sec:experimentbypass}

\textbf{Setup.} We proposed two methods that can bypass the current verification schemes in Section~\ref{sec:bypass}. In this Section, we will evaluate whether our verification scheme remains effective in the presence of these bypass methods. For comparison, we also evaluate both backdoor-based and MIAs-based verification schemes~\cite{TIFSYu2024,DBLP:journals/popets/SommerSWM22}:

\begin{itemize}
    \item For MIAs-based verification schemes, we attempt to bypass the scheme in~\cite{DBLP:conf/uss/LiuWH000CF022}. We select three training samples as member samples to mimic those requiring unlearning. Then, we choose one nonmember sample from the test dataset with the same label as these three samples. Following this, we fine-tune the model using Equation~\ref{equation:attack_loss}, where $\lambda$ is set to $0.01$. The model outputs of the samples before and after fine-tuning are shown in Figure~\ref{fig:attackmia_outputs_before} and Figure~\ref{fig:attackmia_outputs_after}, respectively. The x-axis denotes the index of each posterior probability, and the y-axis indicates the corresponding value. The results of the MIAs are presented in Figure~\ref{fig:attackmia_results}.
    \item For backdoor-based verification schemes, we conduct our experiments based on the scheme in~\cite{TIFSYu2024}. Then, we fine-tuned the model based on Equation~\ref{equation:attack_loss_for_backdoor_random}. We record both the model's overall accuracy and the accuracy of the backdoor. The results are shown in Figure~\ref{fig:backdoor_results}.
    \item  For our scheme, for Equation~\ref{equation:attack_loss}, we employ the same settings as MIAs-based verification schemes. Also, in Figures~\ref{fig:attackour_outputs_before} and ~\ref{fig:attackour_outputs_after}, the x-axis denotes the index of each posterior probability, and the y-axis indicates the corresponding value. For Equation~\ref{equation:attack_loss_for_backdoor_random}, since we maintain the original labels for all samples without alteration, we use these original labels in the fine-tuning procedure. In both processes, we record the same metrics as MIAs-based and backdoor-based verification schemes. The results are shown in Figure~\ref{fig:Bypassing_Schemes_for_our}.
\end{itemize}

\textbf{Results.} As shown in Figure~\ref{fig:attackmia_outputs_before}, before fine-tuning the model, the outputs for member samples and nonmember samples are different, enabling MIAs to verify that samples have not been unlearned based on the model's output. In Figure~\ref{fig:attackmia_outputs_after}, the fine-tuning process makes the output of member samples nearly indistinguishable from that of nonmember samples. This makes MIAs unable to determine whether member samples are used to train the model. As illustrated in Figure~\ref{fig:attackmia_results}, as fine-tuning progresses, the accuracy of MIAs in identifying member samples gradually decreases until they can no longer identify them. This will result in verification schemes based on MIAs being unable to verify whether these samples have been unlearned. Meanwhile, the model's accuracy remains nearly unchanged, indicating that the model provider has successfully bypassed the verification schemes. 

As shown in Figure~\ref{fig:backdoor_results} for backdoor-based verification schemes, after the training process, if a sample contains a backdoor, the model can correctly classify that sample as a backdoored sample. In this case, a backdoor-based verification scheme is effective, meaning that once the backdoor samples are unlearned, the model will no longer be able to classify backdoored samples into a specific target. However, when fine-tuning the model, it cannot correctly classify backdoored samples. This results in the backdoor-based verification scheme becoming ineffective as, even if the model provider has not unlearned the backdoor samples, the model may still fail to correctly classify samples containing a backdoor.

For IndirectVerify, as shown in Figure~\ref{fig:attackour_outputs_before} and Figure~\ref{fig:attackour_outputs_after}, even if the outputs align closely with the output of one test sample after fine-tuning, our verification results show no change~(as shown in Figure~\ref{fig:ourmia_results}), which indicates that IndirectVerify can effectively resist fine-tuning bypass schemes based on Equation~\ref{equation:attack_loss}. This is because IndirectVerify incorporates two samples: one used for unlearning requests and the other for verification. Changing one sample’s output does not affect the classification result of another sample.
Similarly, for the bypass scheme based on Equation~\ref{equation:attack_loss_for_backdoor_random}, as shown in Figure~\ref{fig:ourbackdoor_results}, IndirectVerify also remains effective. Overall, IndirectVerify demonstrates effective resistance against currently known bypass schemes and exhibits a certain level of robustness.

\subsection{Evaluating the Harmlessness}

\textbf{Setup.} To evaluate if the embedded trigger samples affect the model's performance, we evaluate our schemes using the VGG and ResNet models on two datasets: CIFAR-10 and SVHN, respectively. We perturb $1\%$, $2\%$, $3\%$, and $4\%$ of the training samples to trigger samples, respectively, for later verification purposes. We record the model's accuracy before and after embedding the trigger samples. The experimental results are shown in Figure~\ref{fig:modelaccuracy}.

\textbf{Results.} As can be seen from all the figures, the embedding of the trigger samples has an almost negligible effect on the model performance. For example, when considering the ResNet model and the CIFAR-10 dataset, the difference before and after embedding is less than $0.01\%$. This small difference can be attributed to the randomness inherent in the training process. The reason why the impact on the model is minimal is that we did not change the labels of the samples used for verification; we only modified the pixel values of those samples. Furthermore, these modifications are almost imperceptible. Figure~\ref{fig:verificationimage} illustrates a few of the original images and the corresponding trigger samples. It is noted that the two sets of images are almost the same, with no significant difference.

\section{Conclusion and Future Work}
\label{sec:conclusion}
In this paper, we take the first step in addressing machine unlearning verification problems when faced with untrustworthy model providers in MLaaS. We first propose two distinct bypass schemes to illustrate the shortcomings of current membership inference and backdoor attack-based verification methods. Then, we propose a perturbation-based verification scheme to generate influential sample pairs, including trigger samples and a reaction sample. Trigger samples are utilized in the unlearning request, while reaction samples are used for subsequent verification, with the presence or absence of trigger samples influencing reaction samples. We also provide theoretical proofs and analyses of the effectiveness of the proposed scheme, and briefly emphasize its robustness by explaining how our scheme resists current bypassing methods. The experimental results provide evidence that our scheme effectively maintains model utility while achieving robust verification effectiveness.


\bibliographystyle{IEEEtran}
\bibliography{sampleBibFile}
\end{document}